\documentclass{emulateapj}
\usepackage{natbib}
\usepackage{longtable}
\usepackage{rotating}
\usepackage{threeparttable}

\usepackage{amsmath}
\usepackage{subfigure}
\shorttitle{A census of galaxy constituents in a Coma progenitor}
\shortauthors{Shi et al.}

\begin{document}

\title{A census of galaxy constituents in a Coma Progenitor observed at $z>3$}

%\correspondingauthor{Ke Shi}
%\email{shi185@purdue.edu}

\author{Ke Shi\altaffilmark{1}, Kyoung-Soo Lee\altaffilmark{1},   Arjun Dey\altaffilmark{2},  Yun Huang\altaffilmark{1}, Nicola Malavasi\altaffilmark{1,3}, Chao-Ling Hung\altaffilmark{4}, Hanae Inami\altaffilmark{5}, Matthew Ashby\altaffilmark{6}, Kenneth Duncan\altaffilmark{7}, Rui Xue\altaffilmark{1,8}, Naveen Reddy\altaffilmark{9}, Sungryong Hong\altaffilmark{10}, Buell T. Jannuzi\altaffilmark{11},  Michael C. Cooper\altaffilmark{12},  Anthony H. Gonzalez\altaffilmark{13}, Huub J. A. R\"{o}ttgering\altaffilmark{7}, Phillip N. Best\altaffilmark{14}, Cyril Tasse\altaffilmark{15}}
\altaffiltext{1}{Department of Physics and Astronomy, Purdue University, 525 Northwestern Avenue, West Lafayette, IN 47907}
\altaffiltext{2}{National Optical Astronomy Observatory, Tucson, AZ 85726}
\altaffiltext{3}{Institut d'Astrophysique Spatiale, CNRS (UMR 8617), Universit$\acute{\rm e}$ Paris-Sud, B$\hat{\rm a}$timent 121, Orsay, France }
\altaffiltext{4}{Department of Physics, Manhattan College, 4513 Manhattan College Parkway, Riverdale, NY 10471 }
\altaffiltext{5}{Observatoire de Lyon, 9 avenue Charles Andre, Saint-Genis Laval Cedex F-69561, France}
\altaffiltext{6}{Harvard-Smithsonian Center for Astrophysics, 60 Garden St., Cambridge, MA 02138}
\altaffiltext{7}{Leiden Observatory, Leiden University, NL-2300 RA Leiden, the Netherlands 0000-0001-6889-8388}
\altaffiltext{8}{Department of Physics \& Astronomy, The University of Iowa, 203 Van Allen Hall, Iowa City, IA 52242, USA}
\altaffiltext{9}{Department of Physics and Astronomy, University of California, Riverside, 900 University Avenue, Riverside, CA 92521, USA}
\altaffiltext{10}{School of Physics, Korea Institute for Advanced Study, 85 Hoegiro, Dongdaemun-gu, Seoul 02455, Republic of Korea}
\altaffiltext{11}{Steward Observatory, University of Arizona, Tucson, AZ 85721}
\altaffiltext{12}{Department of Physics and Astronomy, University of California, Irvine, CA 92697, USA}
\altaffiltext{13}{Department of Astronomy, University of Florida, Gainesville, FL 32611}
\altaffiltext{14}{Institute for Astronomy, Royal Observatory, Blackford Hill, Edinburgh EH9 3HJ, UK}
\altaffiltext{15}{Observatoire de Paris, CNRS, Universite Paris Diderot, 5 place Jules Janssen, 92190 Meudon, France} 
\begin{abstract}

We present a  detailed census of galaxies in and around  PC217.96+32.3, a spectroscopically confirmed Coma analog at $z=3.78$.  Diverse galaxy types  identified in the field include  Ly$\alpha$ emitters (LAEs), massive  star-forming galaxies, and ultra-massive galaxies ($\gtrsim 10^{11}M_\odot$) which may have already halted their star formation.
The sky distribution of the star-forming galaxies  suggests the presence of a significant  overdensity ($\delta_{\rm SFG}\approx 8\pm2$), which is spatially offset from the previously confirmed  members by 3--4~Mpc to the west. Candidate quiescent and post-starburst galaxies are also found in large excess  (a factor of $\sim$8--15 higher surface density than the field) although their redshifts are less certain. We estimate that the total enclosed mass traced by candidate star-forming galaxies is roughly comparable to that of PC217.96+32.3 traced by the LAEs. We speculate that the true extent of P217.96+32.3 may be  larger than previously known, a half of which is missed by our LAE selection. Alternatively,  the newly discovered overdensity may belong to another Coma progenitor not  associated with PC217.96+32.3. Expectations from theory suggest that both scenarios are equally unlikely ($<1$\%), in the cosmic volume probed in our survey.  If confirmed as a single structure, its total mass will be well in excess of Coma, making this an exceptionally large cosmic structure rarely seen even in large cosmological simulations. Finally, we find that the protocluster galaxies follow the same SFR-$M_{*}$ scaling relation as the field galaxies, suggesting that the environmental effect  at $z\sim4$ is a subtle one at best for normal star-forming galaxies.

\end{abstract}


%\section{notes}

\section{Introduction} \label{sec:intro}

Local environment has a profound influence on the formation and evolution of galaxies. At low redshift, galaxies in dense cluster environments tend to be more massive,  contain older stellar populations,  have lower star formation rates and dust content, and a higher fraction have elliptical morphologies than their average field counterparts \citep[][]{stanford98, blakeslee03,vandokkum07,eisenhardt08}. The redshift evolution of the  cluster red sequence and the properties of cluster ellipticals strongly support a scenario in which cluster galaxies underwent early accelerated formation followed by swift quenching \citep[e.g.,][]{thomas05,Bolzonella10,mancone10,Fritz14}. While this general picture is accepted, the mechanisms responsible for the formation, evolution and quenching processes are still not well understood \citep[e.g.,][]{snyder12}.

In high-density environments, the accretion rates of infalling gas and the frequency of galaxy interactions are expected to be higher, fostering enhanced star formation activities. A merger of gas-rich galaxies may include an ultra-luminous infrared galaxy \citep[ULIRG;][]{Aaronson84,sanders96} phase which efficiently converts the majority of their gas into stars over a short timescale. Dissipative gas-rich mergers may help the efficient feeding of gas into the central blackholes, triggering nuclear activity, which may quench star formation and create old, massive cluster ellipticals. \citep{hopkins08}. High-density environments are therefore expected to consist of diverse galaxy constituents, including normal star-forming galaxies, ULIRGs, X-ray sources, AGN, and massive quiescent galaxies. A detailed census of diverse galaxy `types' and their spatial distribution within the large-scale structure are essential to obtain a more comprehensive understanding of how the high density environment drives the evolution.

To directly witness the key epoch of cluster galaxy formation,  one needs to identify the galaxy populations residing in young `protoclusters'. In recent years, substantial progress has been made in the search for high-z protoclusters \citep[see review by][and references therein]{overzier16}. Searches around powerful radio sources at high redshift have identified significant galaxy overdensities \citep[e.g.,][]{venemans05, miley06, kajisawa06, venemans07,hatch11a,Noirot18}. A population of extremely dusty starburst systems, optically or X-ray-luminous AGN, and large Ly$\alpha$ nebulae are reported in some of the known protoclusters \citep{matsuda04,dey05, prescott08, lehmer09, capak11,casey16,hung16,cai17,badescu17}, in support of the theoretical expectations \citep[but see][]{rigby14,kato16}. The existence of massive `red and dead' galaxy candidates at $z\sim3$  offers tantalizing evidence that the formation of massive cluster ellipticals  may have been well underway as early as 2 Gyr after the Big Bang \citep{kubo13}. 

The number of confirmed protoclusters and protocluster candidates has been increasing rapidly \citep[e.g.,][]{lemaux14, cucciati14, toshikawa16, planck_overdensity,lemaux17,cucciati18}, 
offering a promising outlook for future protocluster studies;  such as the impact of environment on the galaxy inhabitants, as well as the evolutionary link between unvirialized proto-structures and present-day clusters. 

Despite this progress, a clear and coherent physical picture of how cluster environment influences galaxy formation has yet to emerge. We do not yet know how dense protocluster environments influence the galaxy therein: e.g.,  are rare systems such as radio galaxies, quasars, Ly$\alpha$ nebulae ubiquitous enough to be used as beacons of the highest density peaks of the universe?  Do dense protocluster environments produce a different `zoo' of galaxy constituents, or simply a scaled-up version of the average field?  Addressing such questions may have an important cosmological implication: given their large pre-virialization volume and high galaxy overdensities, star formation in protoclusters can account for up to 30\% of the cosmic star formation rate density at $z=4$ \citep{chiang17}.

Observationally,  one of the main limitations has been the lack of our knowledge of the density structure of protoclusters. The angular size of the cosmic volume that will end up virialized by the present-day epoch is expected to be as large as 20\arcmin\ -- 30\arcmin\ in the sky \citep[e.g.,][]{chiang13,muldrew15}, making it  expensive to rely on blind spectroscopic programs to map out their structures with reasonable precision. To date, only a few systems exist with a detailed characterization of their sizes and density structures   \citep[][]{matsuda05,lee14,badescu17}.

Another critical element in making  progress is to obtain a detailed census of protocluster constituents. Understanding how different types of galaxy constituents are distributed within the large-scale structure is necessary to make a fair assessment of how the formation of galaxies is impacted by the environment in which they reside. For example, luminous Ly$\alpha$ nebulae are often found located at the outskirts or an intersection of the densest regions of a protocluster \citep{matsuda05,badescu17}. Several studies reported that powerful AGN  may suppress low-level star formation activity and produce a deficit of Ly$\alpha$-emitting galaxies \citep[e.g.,][]{kashikawa07,goto17} although claims to the contrary also exist \citep[e.g.,][]{cai18}. 

\begin{figure}[h!]
\epsscale{1.2}
\plotone{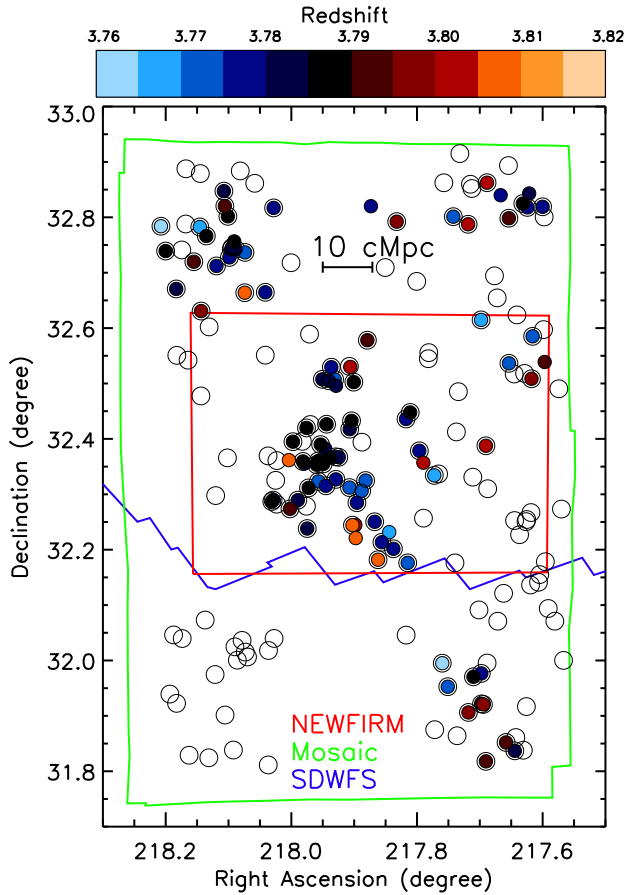}
\caption{
The layout of our protocluster survey field is shown for the Mosaic ($B_WRI$: green), NEWFIRM ($HK_S$: red), and SDWFS  data (blue to the north). The Subaru  $y$-band data covers the field shown here in its entirety. Open circles denote the positions of photometrically selected LAEs, while filled circles show the spectroscopic sources in the range $z$=3.76--3.82, color coded by the redshift indicated by bar on top.  PC217.96+32.3 is situated in the middle of our Mosaic field. 
}
\label{field_layout}
\end{figure}

In this paper, we present a multi-wavelength study of galaxies along the sightline to the PC217.96+32.3 protocluster at $z=3.78$, one of the most massive protoclusters discovered to date \citep{lee14}. Existing spectroscopy has confirmed 48 members at $z$=3.76--3.81 \citep[of which 34 lie at $z$=3.77--3.79;][]{dey16}. The locations of these members are indicated in Figure~\ref{field_layout}. The three-dimensional `map' of the spectroscopic members suggests that the structure is mainly composed of two large groups with a small velocity offset and of additional smaller groups falling in toward the center \citep{dey16}. Given the level and angular extent of the galaxy overdensity, PC217.96+32.3 will likely collapse into a system with  a present-day mass of $M_{\rm total}\gtrsim 10^{15}{\rm M}_{\odot}$, making it one of the few spectroscopically confirmed Coma progenitors. 

Having established the significance of the structure, we are motivated to take a broader view of the constituents of PC217.96+32.3; in particular, we are interested in identifying  more evolved galaxies which may be more closely linked to massive cluster ellipticals in the present-day universe. To this end, we have conducted a deep near-infrared imaging survey of the region, sampling the  continuum emission at rest-frame visible wavelengths. 

In this paper, we present new near-infrared $H$ and $K_S$-band imaging on the central portion of the protocluster field (\S2). Combining this with existing optical data from the NOAO Deep Wide-Field survey  \citep[NDWFS:][]{ndwfs} and mid-infrared data from the {\it Spitzer Space Telescope} \citep{ashby09}, we identify a large overdensity of luminous galaxies in the region (\S3). Population synthesis modeling of these galaxies suggests that they are likely to lie close to the redshift of the protocluster traced by the Lyman Alpha emitters (LAEs), although they have a somewhat different spatial distribution (\S4). We discuss the masses, star-formation rates, and estimate the size of the overdensity in \S4, and discuss the implications of finding such an overdense region in \S5. 

Throughout this paper, we use the WMAP7 cosmology $(\Omega_m, \Omega_\Lambda, \sigma_8, h)=(0.27, 0.73, 0.8, 0.7)$ from \citet{wmap7}. Distance scales are given in comoving units unless noted otherwise. Magnitudes are given in the AB system \citep{oke83} unless noted otherwise. In the adopted cosmology, PC217.96+32.3 at $z=3.78$ is observed when the universe was 1.7~Gyr old; 1\arcmin\ corresponds to the physical scale of 2.1~Mpc at this redshift.

\section{Data and Photometry} \label{sec:data}

The  multi-wavelength data available in this field include the optical data taken with three broad-band filters  ($B_WRI$: NOAO program IDs: 2012A-0454, 2014A-0164) using the Mosaic camera \citep{jacoby98,mosaic3} on the Mayall telescope, and the {\it Spitzer} IRAC 3.6$\mu$m, 4.5$\mu$m, 5.8$\mu$m, and 8.0$\mu$m data taken as part of the Spitzer Deep Wide-Field Survey \citep[SDWFS:][]{ashby09}. As discussed in \citet{dey16}, the new optical $B_WRI$ data are combined with the reprocessed NDWFS data \citep{ndwfs} to create the final mosaicked images. 

We obtained $y$-band imaging from Hyper Suprime-Cam \citep[HSC:][]{miyazaki18} on the Subaru telescope, which provides the field-of-view of 1.77 deg$^2$ and the pixel scale of 0\farcs168. The observations were carried out on March 27, 2015 with typical seeing of $\sim$0.6\arcsec, and consisted of 200~sec exposures with the total exposure time of 2.4 hours. The individual images were reduced and coadded using the HSC data processing pipeline \citep{bosch18}. The pipeline performed standard bias, dark, flat, and fringe calibrations, and the astrometry and photometry were calibrated based on Pan-STARRS1 surveys \citep{chambers16} before coadding.

In March 2015 and March 2016, we obtained deep imaging of the survey field using the NEWFIRM camera \citep[][NOAO program IDs: 2015A-0168, 2016A-0185]{newfirm1,newfirm2} on the Mayall 4m telescope of the Kitt Peak National Observatory. The camera has a  pixel size of 0.4\arcsec\ and covers a 28\arcmin$\times$28\arcmin\ field of view. Images were obtained with $H$ and $K_S$ bands (KPNO filter no. HX (k3104) and KXs (k4102); $\lambda_{\rm c}$=16310 and 21500 \AA\ with the full-width-at-half-maximum (FWHM) of 3080 and  3200 \AA\, respectively. We will refer to these filters as $H$ and $K_S$ band, hereafter). The pointing center -- $\alpha$=14:31:28.8, $\delta$=32:23:24.0 (J2000) --  was chosen to cover the known protocluster region  in its entirety while sampling a sufficient flanking region outside of it. We used individual exposure times of 60~sec for both bands, and dithered the telescope between exposures up to 2\arcmin\ in random directions using the {\tt DEEPSPARSE} dither pattern. 

Each science frame is dark-subtracted and flat-fielded using the standard NOAO pipeline. We calibrate the astrometry using stars identified in the Sloan Digital Sky Survey DR7 catalog, and reproject each frame to a common tangent point with a pixel scale of 0$\farcs$258 in order to match that of the  optical data.  The relative intensity scale of each  frame was determined using the {\tt mscimatch} task.  The reprojected images were combined into a final stack using a relative weight inversely proportional to the variance of the sky noise.  Only the frames with the delivered image quality of seeing $\geq 1.3$\arcsec\ are included in the image stack. We trim the image borders whose exposure is less than 20\% of the maximum exposure time, and obtain the final coadded mosaic with an effective area of 28\arcmin$\times$35\arcmin~(0.27~deg$^2$). The effective total exposure times of the mosaics are 12.1 and 18.7 hours for the $H$ and $K_S$ band, respectively. The photometric zeropoints are determined by cross-correlating the detected sources with the 2MASS point source catalog. 

We resample the Spitzer SDWFS data to have a pixel scale of 0\farcs774, i.e., three times larger than the optical/near-IR data. Having the pixel scales to be integer multiples of one another is necessary for extracting optical photometry via a template-fitting method (see later). The $5\sigma$ limiting magnitudes measured in a 2\arcsec\ diameter aperture are 26.88, 26.19, 25.37, and 25.10~AB in the optical data ($B_WRIy$), 24.05 and 24.83~AB in the near-IR data ($HK_S$), respectively. The seeing measured in the stacked images is 1.0\arcsec\ in the $B_WRI$ images, 0.6\arcsec\ in the $y$-band, and 1.2\arcsec\ in the $HK_S$ bands. The sky coverage of our dataset is illustrated in Figure~\ref{field_layout}. 

\subsection{Multi-wavelength Catalog} 

We use the PSFEx software  \citep{psfex} to measure the  point spread function (PSF) of each image out to a radius of 3\arcsec. The two-dimensional PSFs are  radially averaged to obtain the circularized PSF.  Taking the worst-seeing data ($K_S$ band) as the target PSF, we derive the noiseless convolution kernel  for each image using the IDL routine {\tt MAX\_ENTROPY}. We use the full shape of the observed stellar profiles rather than assuming a function form such as Moffat profiles. The details of the PSF matching procedure are given in \citet{xue17}. All optical and near-IR images are convolved with the appropriate kernels to create a set of PSF-matched science images. 

Source detection and photometric measurements in the $B_WRIyHK_S$ bands are carried out running the SExtractor software  \citep{bertina96} in dual mode on the PSF-matched images with the $K_S$ band data as detection band. At the protocluster redshift ($z=3.8$), the  $K_S$ band mainly samples the continuum emission at the rest-frame  $\approx 4400$\AA.
%central wavelength of the $K_S$ band samples $\approx 4400$\AA\ in the rest-frame, and thus the $K_S$ band flux should reflect the rest-frame optical brightness of the sources, which could enable us to estimate the stellar mass. 
The SExtractor parameter MAG\_AUTO is used to estimate the total magnitude, while colors are computed from fluxes within a fixed isophotal area (i.e., FLUX\_ISO). Colors measured in FLUX\_ISO and FLUX\_APER are in agreement with each other within 0.1~mag. As the images are PSF-matched, aperture correction is constant in all bands, and is given by the difference between MAG\_AUTO and MAG\_ISO estimated in the $K_S$ band. 

For the {\it Spitzer} IRAC images, we take a different approach as it is not practical to convolve all images to the FWHM of any IRAC PSF, which is much broader ($\sim 2 \arcsec$). We use the TPHOT software \citep{merlin15} which performs `template fitting photometry' similar to TFIT \citep{laidler07, lee12a}. The software uses the information (source shape and position) supplied by a higher-resolution data and simultaneously fits the fluxes of multiple nearby sources to minimize residual flux. Since the FWHM of the $K_S$ band PSF is not negligible compared to that of the IRAC data, we also derive the convolution kernel using the same procedure above. For the effective PSF of the IRAC bands, we rotate the published IRAC PSF by a series of position angles with which the SDWFS data were taken, and create a weighted average image. 

Finally, all photometric catalogs are merged together to create the final multi-wavelength catalog, where the TPHOT-measured fluxes are considered identical to the MAG\_ISO fluxes of the optical/near-IR bands. Given the completeness of the $K_S$ band data, we only consider sources that have the signal-to-noise ratio (SNR) greater than 10, roughly corresponding to $K_{S, {\rm AB}}$ magnitude of 24.0 mag. The final multi-wavelength catalog contains 27,845 sources. 

\subsection{Photometric Redshift and SED Modeling}\label{photo_z}
We derive photometric redshifts with the CIGALE code \citep{noll09} using the full photometric information. The reliability of the photo-$z$ estimates is evaluated using the existing spectroscopic sources, which targeted a subset of UV-bright galaxies satisfying the Lyman Alpha Emitter (LAE) or Lyman Break Galaxy (LBG) color selection over a $1.2\times 0.6$ deg$^2$ contiguous region  in the  PC217.96+32.3 field. The details of these selection methods in our survey field are discussed in \citet{lee13} and \citet{lee14}. Of the 164 sources at $z_{\rm spec}=3.4-4.2$, 48 galaxies lie within the NEWFIRM coverage. Of those, only 17 galaxies are bright enough to be detected in the $K_S$ band catalog with a photo-$z$ estimate. 

We find the redshift dispersion $\sigma_z/(1+z_{\rm spec})$= 0.15 where $\sigma_z$ is the standard deviation of $\Delta z$ ($\equiv z_{\rm spec}-z_{\rm phot}$). The large dispersion is  due to three outliers which have $(z_{\rm{spec}}-z_{\rm phot})/(1+z_{\rm spec}) > 0.2$. We  find that their redshift probability density functions have two peaks, one at $z<1$ and the other at $z\sim 4$; given that they are fainter than other galaxies,  redshift degeneracy is caused by the fact that the spectral break between $B_W$ and $R$ is not strong enough to be unambiguously determined as a Lyman break. However, for all three galaxies the probability to lie at $z=3.4-4.2$ (computed by integrating the photometric redshift probability density function in the interval, which we denote as $p_z$) is greater than 50\%.  
Excluding these three galaxies, the redshift dispersion $\sigma_z/(1+z)$ is 0.06.

After considering the photometric redshift constraints of our spectroscopic sources, we select protocluster candidate galaxies by requiring that  $z_{\rm phot}=3.4-4.2$ for the sources whose redshift probability density functions (PDFs) are singly peaked, and  $p_z \geq 0.5$ for those with doubly peaked PDFs. All of the 17 spectroscopic members meet these criteria. The range $z_{\rm phot}=3.4-4.2$ is chosen based on the photometric redshift error as discussed previously. A similarly inclusive range was used by \citet{kubo13}, who studied the stellar populations in and around another protocluster. %This redshift range is also consistent with the spec-$z$ LBGs in the field and later we will see that the majority of the photo-$z$ candidates are LBG-like galaxies. 
 After visual inspection, we remove the sources with potential contamination in the photometry including those that are too close to brighter sources or to the edges of the images. Our protocluster galaxy sample consists of 263 sources, which also includes the spectroscopic members of the structure.

\begin{figure}[h]
\epsscale{1.1 }
\plotone{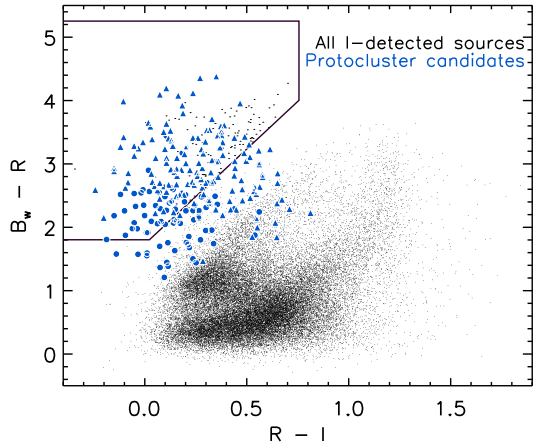}
\caption{
The locations of the photo-$z$ protocluster candidates on the $B_W-R$ vs $R-I$ diagram are shown together with all $I$-band detected sources (black dots). Galaxies that are undetected in the $B_W$ band are shown as upward triangles. The formal LBG criteria to select galaxies at $z\sim3.4-4.2$ are shown as polygon in the upper left corner \citep{lee14}. The majority of our photo-$z$ candidates would formally meet the LBG selection.  }
\label{two_color}
\end{figure}

We examine the rest-frame UV colors of our protocluster candidates to assess their overall similarities to broad-band color-selected LBGs. We match their positions to the $I$-band-selected photometric catalog used for the $B_W$ band dropout  selection \citep{lee13,lee14}. Of the 263 sources, 202 galaxies (77\%) are detected in the $I$-band with the signal-to-noise ratio (SNR) $\geq 7$. In Figure~\ref{two_color}, we show their locations on the $B_W-R$ vs $R-I$ color diagram. Of the 202 galaxies, 135 galaxies (67\%) satisfy the formal LBG criteria, and an additional 22 galaxies (11\%) are within 0.25~mag of the formal $B_W-R$ color cut. The galaxies outside the selection window tend to be fainter in the $R$ and $I$ bands ($R\lesssim 25.5$) while dropping out of the $B_W$ band, which results in a weaker constraint on the $B_W-R$ color. However, their  UV colors are generally similar to their UV-brighter cousins. Their $R-I$ colors are  redder than those within the LBG selection criteria, which likely contribute to a weaker spectral break in the $B_W$ band. Thus, we conclude that our photo-$z$ estimate works relatively well for moderately dust obscured star-forming galaxies whose spectral energy distributions (SEDs) are similar to those of LBGs.

\begin{figure*}[ht!]
\epsscale{1.2}
\plotone{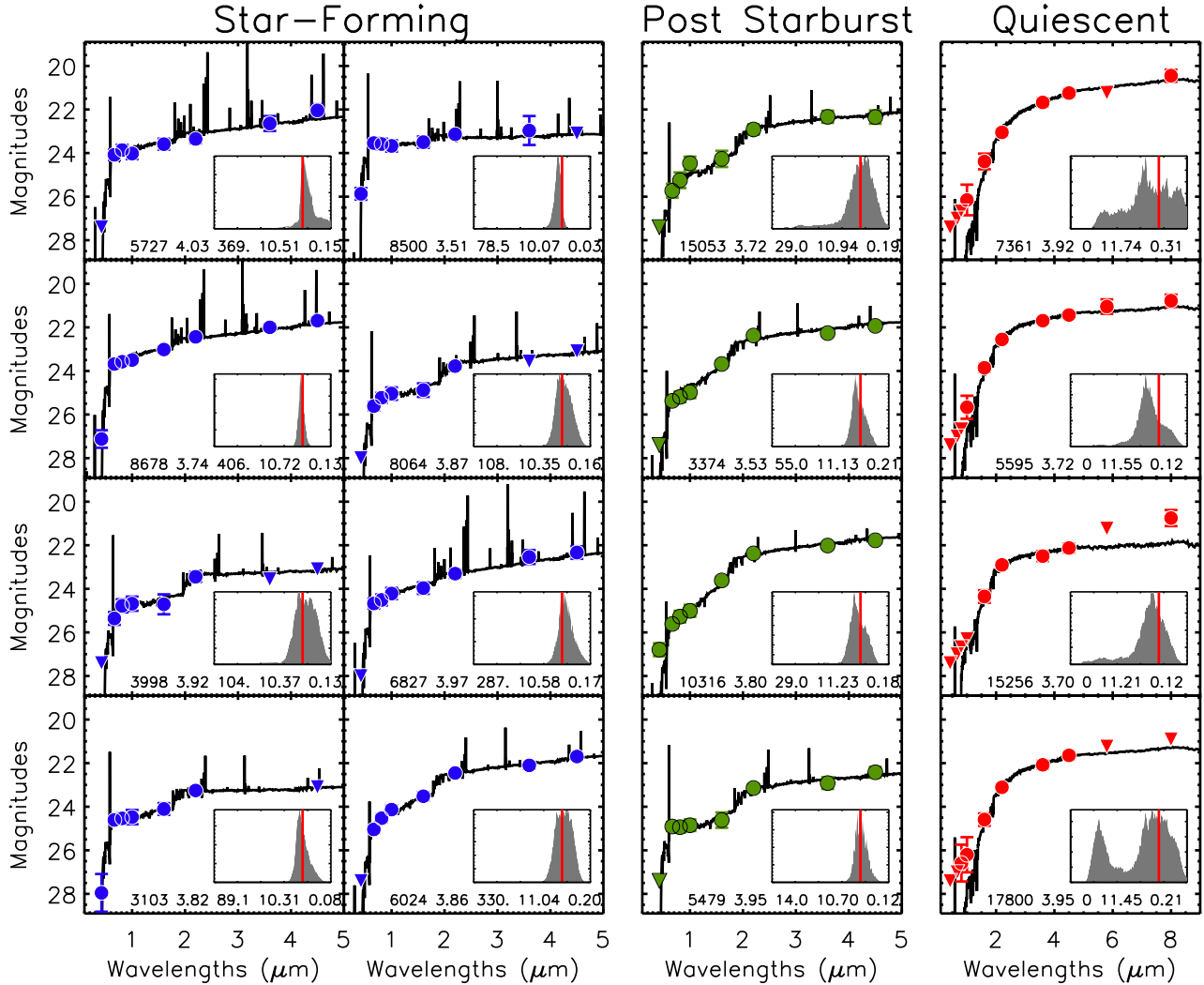}
\caption{
Observed SEDs are shown for a subset of our galaxy candidates, which include normal UV-bright star-forming galaxies (blue), post-starburst candidates (green), and quiescent galaxy candidates (red). Filled circles represent our photometric measurements, while triangles denote $2\sigma$ limits in the case of non-detection. We also show the CIGALE redshift probability density functions as shaded grey regions (inset); the plotting range is $z=[0,5]$. The redshift of PC217.96+32.3 is shown as a red vertical line. On bottom of each subpanel, we list object ID, best-fit photo-$z$, star formation rates (in units of $M_\odot~{\rm yr}^{-1}$), $\log{(M_{\rm star})}$ (in units of $M_\odot$), and dust reddening parameter E($B-V$).
}
\label{sed_montage}
\end{figure*}

We determine the physical properties of the photo-$z$ protocluster candidates using the CIGALE software. We use the stellar population synthesis models of \citet{bc03}, the \citet{calzetti00} reddening law with $A_V$ values ranging from 0 to 5 in steps of 0.1~mag, solar metallicity, and \citet{Salpeter55} initial mass function. We carry out three separate runs assuming a constant star formation histories, an exponentially declining SFH with $\tau$ values from 50 Myr to 10 Gyr in steps of 200 Myr, and a delayed star formation model. In Figure~\ref{sed_montage} (left two columns), we show the best-fit SED model, population parameters, and redshift PDF (inset) for a subset of our sample. As can be seen in the figure, these photo-$z$ candidates are mostly star-forming galaxies with a strong Ly$\alpha$ break and relatively blue UV slope. The best-fit star-formation rate implies they are actively forming stars at a relatively high level (up to several hundreds solar mass per year) with a moderate amount of dust (E($B-V$)$\sim$0.1 -- 0.2).

\section{Balmer-break Galaxy Candidates in the Protocluster Field}\label{sec_bbg}

\subsection{Selection of galaxies with evolved stellar populations}\label{selection_bbg}
As discussed in \S~\ref{photo_z}, the photometric redshift technique is most effective in selecting LBG-like galaxies.  Here, we use  a set of color selection criteria tuned to isolate galaxies with a strong Balmer/4000\AA\ break, a feature strongest in old stellar populations dominated by A and F stars. In this work, we use the following color criteria, which are similar to those found in the literature \citep[e.g.,][]{franx03,labbe05,kajisawa06,brammer07, wiklind08, huang11,nayyeri14, mawatari16}:
\begin{align*}
H-K_S>1.2;\\
[3.6]-[4.5]<0.5
\end{align*}
\begin{figure*}[ht!]
\epsscale{1.15}
\plotone{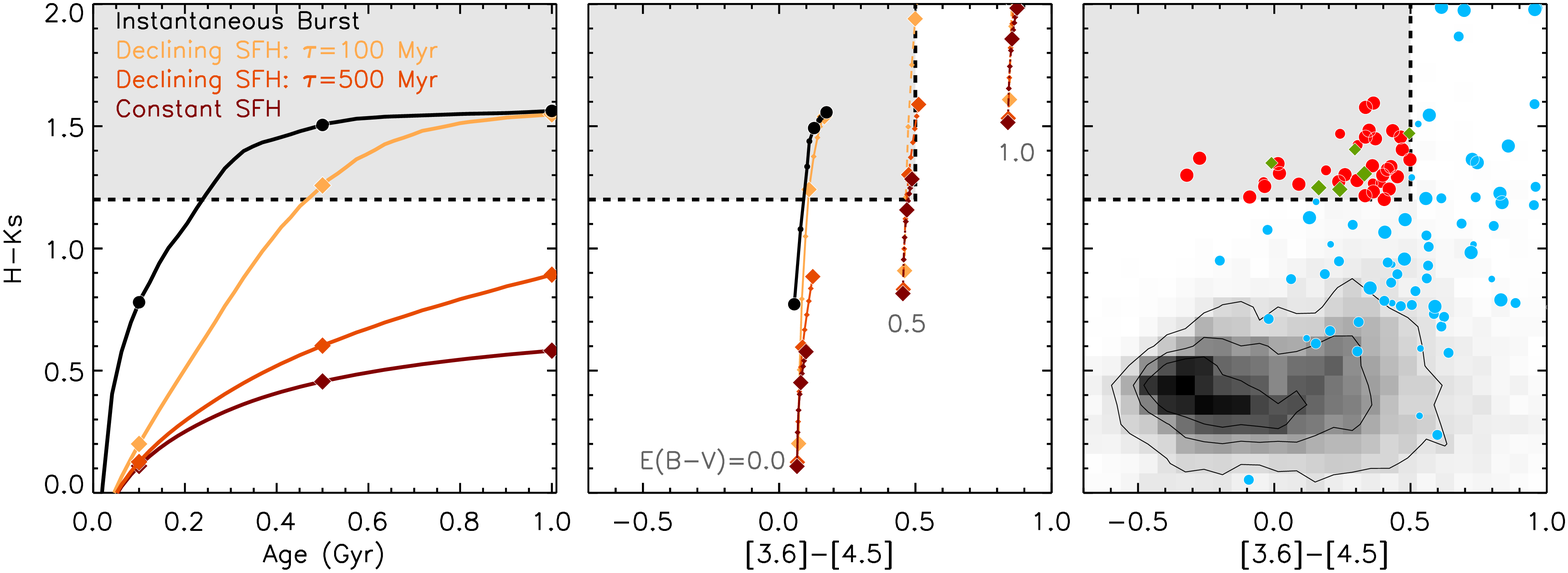}
\caption{
{\it Left:}
expected $H-K_S$ colors are shown as a function of population age for single burst (black), exponentially declining SFH with $\tau=0.1$ Gyr (light orange), $\tau=0.5$ Gyr (dark orange), and constant SFH (brown). Only the galaxies with relative quiescence can achieve the $H-K_S$ color cut (grey shades). 
{\it Middle:} 
the  evolution of $H-K_S$ vs $[3.6]-[4.5]$ colors are shown for three dust reddening parameters  E($B-V$)=0 (solid), 0.5 (dashed), and 1.0 (dotted). Grey shaded region marks our selection criteria for the Balmer/4000\AA\ break galaxies candidates.  The circles in each model mark the population age of 0.1, 0.5, and 1 Gyr, from bottom to top. 
{\it Right:} 
The sources satisfying our BBG criteria are shown in red ($R$-band undetected) and green ($R$-band detected) symbols.  A subset of photo-$z$ protocluster candidates with robust $[3.6]-[4.5]$ color measurements are also indicated (light blue circles).  The grey shades and contours show the distribution of all sources (25\%, 50\%, and 75\% levels). The size of the symbols indicates the stellar masses of the galaxies, classified as $M_\star < 10^{10.5} M_\sun$ (small), $10^{10.5} M_\sun < M_\star < 10^{11} M_\sun$ (medium), and $M_\star > 10^{11} M_\sun$ (large).}
\label{BBG_selection}
\end{figure*}

The first condition imposes that a strong Balmer/4000\AA\ break falls between the $H$ and $K_S$ bands, which occurs in the redshift range  $z=3.6-4.2$. Using the EZGAL software\footnote{{\tt http://www.baryons.org/ezgal/} } \citep{Mancone12} with the stellar population synthesis models of \citet{bc03}, we compute the $H-K_S$ colors of stellar population as a function of age, assuming three families of star formation histories: 1) instantaneous burst; 2) constant star formation histories (CSF); and 3) exponentially declining $\tau$ model (EXP models hereafter: ${\rm SFR}\propto \exp{[-t/\tau]}$) with $\tau$ values of 0.1~Gyr and 0.5~Gyr.  As illustrated in the left panel of Figure~\ref{BBG_selection}, a stellar population formed via a single instantaneous burst would meet this condition at age 250~Myr, while galaxies formed through a more extended star formation episode ($\tau$=100~Myr) would take $\approx$400~Myr to attain the same strength.   

The second criterion requires that the continuum slope at $\lambda_{\rm rest}$ =7000--9000\AA\ is relatively flat,  ensuring that the red $H-K_S$ color is not due to dust reddening. In the middle panel of Figure~\ref{BBG_selection}, we illustrate the effect of interstellar dust assuming the reddening parameters  E($B-V$)=0, 0.5, 1.0 and the \citet{calzetti00} extinction law. 

Using the above criteria, 56 galaxies are identified. Thirteen of them have power-law-like SEDs in the mid-infrared with typical brightness of $\approx 21$~AB in the 5.8$\mu$m or 8.0$\mu$m bands; the $H-K_S$ colors range in $1.2-1.3$, on the low end of the color distribution. Four of them are also present in the {\it Spitzer} MIPS $24\mu$m source catalog provided by \citet{vaccari15}. These sources are likely heavily dust-obscured AGN which scatter into our selection. Of the thirteen galaxies, 7 (54\%) and 8 (62\%) of them meet the IRAC color criteria for high-redshift AGN selection proposed by \citet{stern05} and \citet{donley12}, respectively. We remove all thirteen galaxies from our sample. 

The final  sample consists of 43 galaxies, which we refer to as Balmer break galaxy candidates (BBGs) hereafter. Seven galaxies are also our photo-$z$ protocluster candidates. In the right-most panel of Figure~\ref{BBG_selection}, we show the $H-K_S$ and $[3.6]-[4.5]$ colors of all $K_S$-band detected sources with reliable color measurements. The BBGs without (with) the photo-$z$ estimate are shown in red (green), while the distribution of the remainder is indicated as greyscale and contours where the contour lines enclose the 68\% and 95\% of all galaxies. 

\begin{figure}[h]
\epsscale{1.2}
\plotone{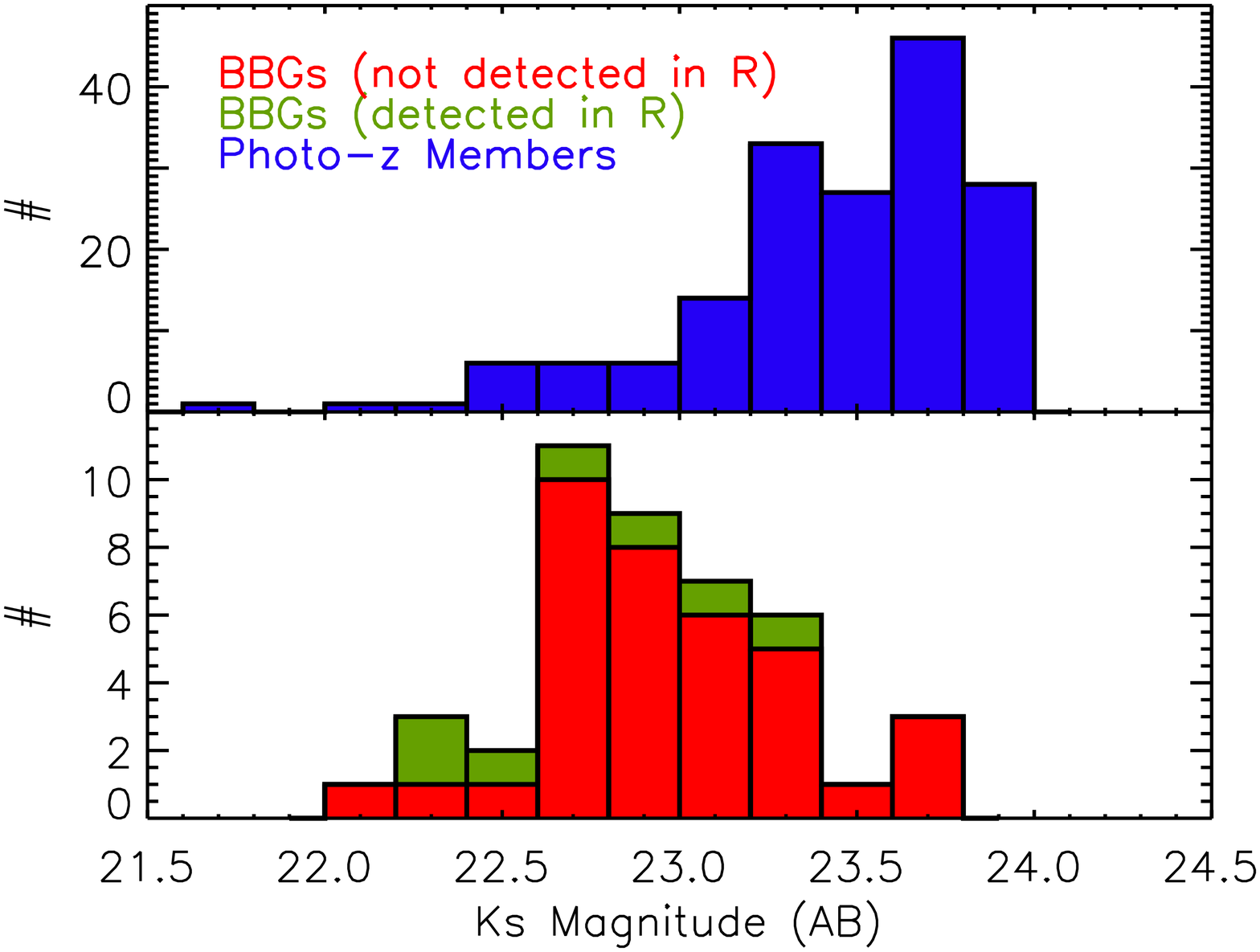}
\caption{
Distribution of $K_S$-band magnitudes is shown for photo-$z$ selected star-forming galaxies (top, blue histogram) and BBGs (bottom). As for the latter, those with and without $R$-band detection are indicated as green and red, respectively. The 2$\sigma$ $R$ band limiting magnitude is 27.2~AB.  The $R$-band samples $\lambda_{\rm rest}\geq 1200$\AA\  at $z\sim3.8$. 
}
\label{maghist}
\end{figure}

Most BBGs are very faint at observed optical  wavelengths (i.e., faint at rest-frame UV wavelengths). Of the 43 galaxies,  only seven (16\%) are detected  at the $5\sigma$ level ($R\leq 26.2$~AB) in the $R$-band while dropping out of the $B_W$ band. The three brightest galaxies (in the $R$-band) formally meet the LBG color criteria. The remaining four likely have similar SED shapes to their $R$-brighter counterparts but are simply too faint  to place strong enough constraints on the $B_W-R$ colors. All seven have photo-$z$ probability distributions with a single peak at $z>3.5$. The remaining 36 galaxies (84\%) are formally undetected in the $R$ band with a few detected at a lower significance. 

In the $K_S$ band, the BBGs have a mean $\langle K_S \rangle=22.94\pm0.37$~AB (median 22.92), significantly brighter than the photo-$z$ members, which have $\langle K_S \rangle=23.44\pm0.40$ (median 23.50). In Figure~\ref{maghist}, we show the $K_S$ band  distribution of the photo-$z$  (solid grey) and BBG candidates (hatched) where the BBGs are further split based on optical detection (labelled as `UV-faint' and `UV-bright'). The disparity between  their $K_S$ band brightness is  driven by a selection effect: the IRAC color cut applied to the latter requires that they have to be bright enough in both IRAC 3.6$\mu$m and 4.5$\mu$m bands. 

We construct the median SED for each BBG subsample by creating median image stacks and measuring aperture photometry \citep[see][for the full description of the stacking procedure]{lee11}.  The CIGALE software is run  in the same manner as done for individual galaxies. The SED fitting results are summarized in Table~\ref{stellar_pop}.
We find that optically faint BBGs are well fit by  old stellar populations with little to no star formation, while the remaining seven galaxies are best-fit as  young post-starburst systems. We refer to the two groups as `quiescent' and `post-starburst' BBGs, respectively. In the next two subsections, we discuss each category in further detail.  

\begin{table*}\label{stellar_pop}
\begin{center}
\caption{Physical properties of BBGs (stacked photometry)}
\begin{tabular}{c|c|ccc}
\hline
\hline
 & Quiescent &   & Post starburst & \\
 &    exp. decl.    & two populations & exp. decl. & CSF \\
SFH &    $\propto \exp{[-t/\tau]}$    & $= C_{\rm old}\exp{[-(t-t_{\rm old})/\tau_{\rm old}]}$  & $\propto \exp{[-t/\tau]}$ & $=const$ \\
   &                            &    + $C_{\rm new}\exp{[-(t-t_{\rm new})/\tau_{\rm new}]}$       &               \\
\hline
$z_{\rm{phot}}$ & $3.58\pm0.37 $& $3.96\pm0.26$ & $3.95\pm0.26$ & $3.95\pm0.26$ \\
$\log{[M_{\rm{star}}/M_\odot}]$ &  $11.20\pm0.07$ & $10.99\pm0.09$ & $10.99\pm0.10$ & $10.95\pm0.09$ \\
SFR ($M_\odot$~yr$^{-1}$)& $0\pm2$ & $114\pm61$ & $110\pm69$ & $172\pm68$ \\
Age (Myr) & $984\pm324$ & $395\pm141$ & $405\pm154$ & $358\pm127$ \\
E($B-V$) & $0.08\pm0.08$ & $0.16\pm0.04$ & $0.16\pm0.05$ & $0.19\pm0.03$ \\
$\tau$ (Myr) & 50 & 100, 300 & 500 & $\infty$ \\
$f_{\rm new}^\dagger$ & -  & $\lesssim 0.1$ & - &  - \\
$\chi_r^2$ &  6.39 & 5.22 & 5.29 & 5.38 \\
\hline
\end{tabular}
\tablecomments{$\dagger$ the fraction of stellar mass formed in the second burst relative to the older population. }
\end{center}
\end{table*}

\begin{figure*}[ht!]
\epsscale{1.0}
\plotone{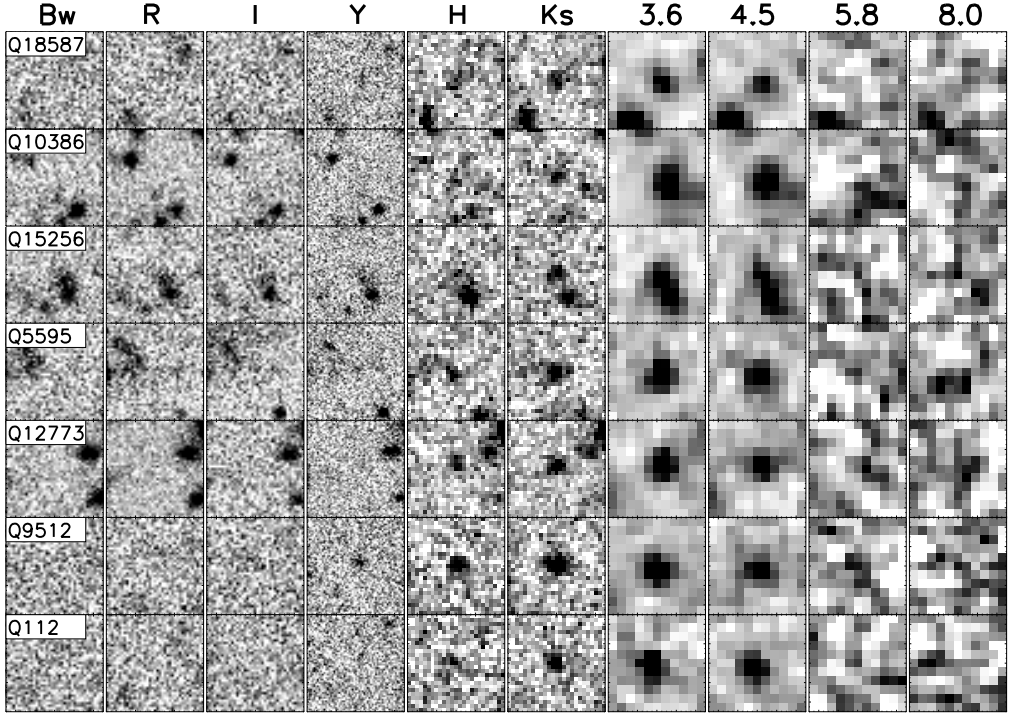}
\caption{
Postage-stamp images of example quiescent BBG candidates. All images are 10\arcsec\ on a side (north is up and east is to the left).
}
\label{stamps}
\end{figure*}

\begin{figure*}[ht!]
\epsscale{1.1}
\plotone{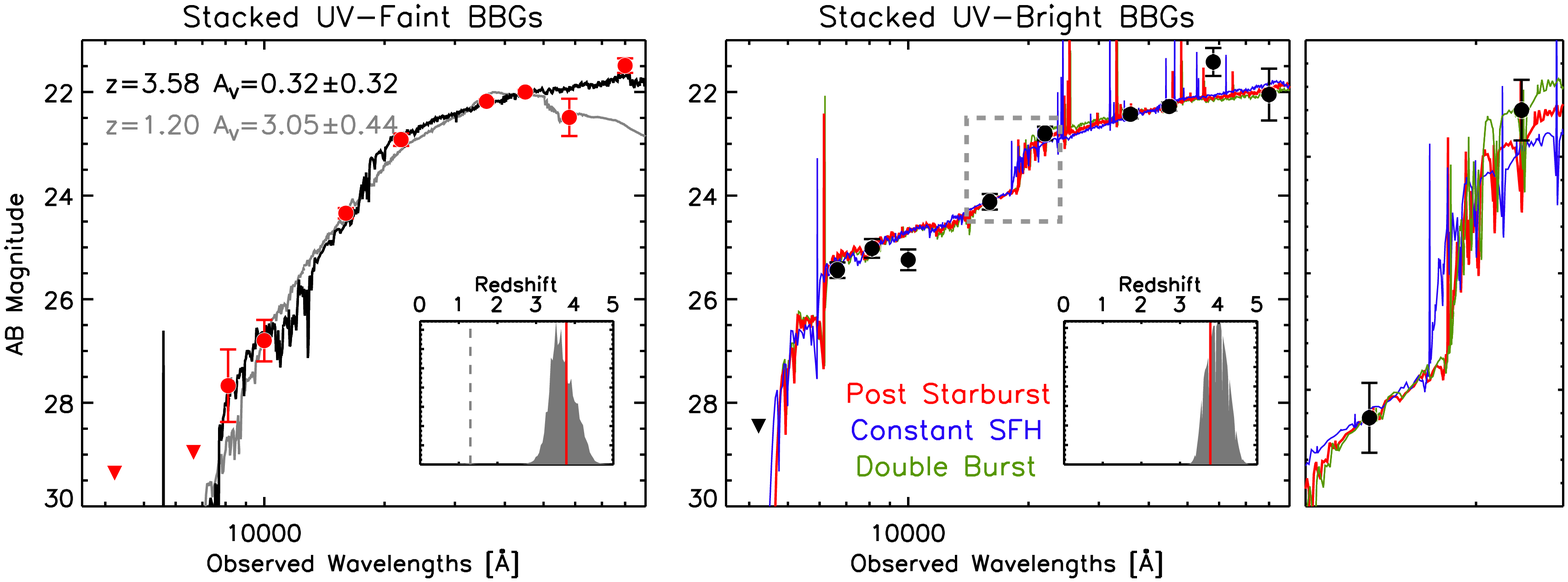}
\caption{
Photometry performed on image stacks created for BBG candidates are shown together with the CIGALE-derived best-fit SEDs. Inset shows the redshift PDF as grey histogram where the redshift of PC217.96+32.3 is marked as vertical red line. {\it Left:} the median-stacked SED of the 36 optically faint BBGs is consistent with that of a very massive and evolved galaxy at $z\sim3.6$ (black). The SED of an old and very dusty galaxy at $z=1.2$ is shown in light grey, highlighting its similarity in optical and IR color to a quiescent galaxy at $z\sim3.6$. However, for the lower redshift ($z<1.5$) solution, a turnover in the grey model falls between 4.5 -- 5.8$\mu$m due to the stellar bump in the rest-frame 1.65$\mu$m. 
{\it Right:} the median-stacked SED of the 7 optically bright BBG candidates is shown with three best-fit models, namely post-starburst (red), constant SFH (blue), and double-burst (green); all three models have very similar SED shapes except for subtle differences near the Balmer/4000\AA\ break. A zoom-in of the region outlined by a grey dashed box is shown on right (see \S\ref{psb}). 
}
\label{stacked_sed}
\end{figure*}

\subsection{Quiescent Galaxy Candidates}\label{bbg_q}

In Figure~\ref{stamps}, we show sample postage stamp images of the quiescent BBGs. Most of the quiescent BBG candidates are detected only in 3 or 4 bands; the limited dynamic range in the wavelength coverage and shallow depths in the IRAC 5.8$\mu$m and 8.0$\mu$m bands result in poorly constrained photometric redshift estimates. While we return to the issue of redshift degeneracy later in this section, we fix the redshift of all quiescent BBGs to $z=3.8$ in deriving their physical parameters, which is motivated by the redshift of the protocluster in the field. Changing the redshift by $\Delta z=\pm0.1$ would result a 5\% change in mass. 

Twelve BBGs show an excess flux in the 5.8$\mu$m and 8.0$\mu$m bands suggesting possible contamination by warm dust emission,  possibly arising from hidden starburst or AGN. When we exclude the 5.8-8.0$\mu$m data from the SED fitting and refit their masses, the change in stellar mass is minimal (6\%). This is consistent with the expectation based on infrared SEDs of high-redshift starburst/AGN systems that the flux contribution by AGN at $\lambda_{\rm rest} \leq 1-2$~$\mu$m is not significant \citep[e.g.,][]{sajina12,kirkpatrick15}. The median value of the individual stellar mass measurements is $\log{[M_{\rm star}/M_\odot]}$=11.30 ($\sigma$=0.29),  consistent with that obtained from the stacked photometry (Table~1). The four most massive galaxies lie in the range $\log{[M_{\rm star}/M_\odot]}$=11.7--11.9 (see Fig~\ref{sed_montage}); if confirmed, their masses already rival some of the brightest cluster galaxies in the local universe.

The stacked SED of the UV-faint BBGs is consistent with an old (980~Myr) and very massive ($\approx 2\times 10^{11}M_\odot$) galaxy  with little star formation. When combined with the best-fit photometric redshift at $z_{\rm phot}=3.58$, the formation redshift is at  $z_{\rm f}\approx 6$. While similarly massive and old galaxies have been reported in the literature \citep{marchesini10,nayyeri14,glazebrook17}, the presence of such massive galaxies in large number may pose a considerable challenge to the hierarchical theory of galaxy formation. %We return to this issue in the discussion section. 

The redshift PDF of the stacked SED (Figure~\ref{stacked_sed}, left inset) is singly peaked at $z\approx 3.6$ strongly ruling out a lower-redshift ($z<3$) solution. This gives us confidence that our quiescent galaxy sample is not dominated by heavily obscured lower-redshift sources. However, the photo-$z$ constraints on individual galaxies are more ambiguous (Figure~\ref{sed_montage}). Only eleven of the thirty six galaxies have a singly peaked PDF at $z>3$; the remainder shows a rather flat $z$-distribution or has two peaks. For the latter, the low-redshift solution typically lies at $z_{\rm phot}=1.0-1.5$ and the high-$z$ solution lies at $z_{\rm phot}=3.5-4.0$. 

The color degeneracy  between an old quiescent galaxy at high redshift and a very dusty galaxy at lower redshift is well known. \citet{dunlop07} reanalyzed the photometric data of a putative massive and quiescent galaxy at $z=6.5$ named HUDF-JD2 \citep{mobasher05}, and showed that a very dusty ($A_V=3.8$) galaxy at $z\sim 1.5-2.5$ is equally likely.  The galaxy was later detected  in the 16~$\mu$m and 22~$\mu$m bands lending   further credence to the lower-z solution \citep[$z\approx 1.7$:][]{chary07}. Similarly, \citet{marchesini10} selected a sample of massive galaxies at  $z=3-4$ using the photometric redshift technique, and noted that  nearly a half are equally well fit by very old and very dusty ($A_V\approx 3$)  galaxies at $z<3$, if such models  are included in the spectral template set. 

In the redshift range to which our BBG selection is sensitive, the strongest constraint comes from the IRAC 5.8$\mu$m and 8.0$\mu$m photometry. For galaxies at $z\lesssim1.5$, these bands sample beyond the 1.65$\mu$m stellar bump which arises from the declining H$^-$ ion opacity in the stellar atmosphere \citep[see, e.g.,][]{sawicki02}. Indeed, when we repeat the SED fitting procedure while limiting the redshift range to $z<2$, the best-fit solution is an old and heavily reddened galaxy at $z\approx1.2$ ($A_V=3.1\pm0.4$, $2.5\pm1.4$~Gyr) which is shown in Figure~\ref{stacked_sed} (left). Given the similarity of the rest-frame UV and optical colors of the two model fits, it is evident that flux measurements in the 5.8$\mu$m and 8.0$\mu$m band are important in breaking the redshift degeneracy. 

Indeed, all of the BBGs with the 8.0$\mu$m detection have  singly peaked redshift PDFs. In the image stack of the remaining 25 galaxies, we do not obtain a clear detection, and as a result, the redshift PDF is doubly peaked confirming our expectation. However, the non-detection is not surprising considering  the sensitivity of the SDWFS data  (5$\sigma$ limit for a point source is 20.25~mag). If we assume  Poisson noise (i.e., the most optimistic case),  stacking 25 sources would result in the limiting magnitude of 22.0, which is very close to the measured 8.0$\mu$m flux from the full stack (see Fig.~7). Thus, the non-detection does not rule out the possibility that these 25 galaxies have similar SEDs to the 8.0$\mu$m-brighter counterparts but with slightly lower fluxes. 

As a final check, we repeat our image stacking, photometry, and SED fitting procedure for 200 times while each time randomly excluding 7 BBGs (20\% of the sample). We integrate the redshift PDF above $z=3$ to obtain the formal probability $P_3$ for the high-redshift solution. In 73\% of the time, the photometric redshift solution prefers the high-z solution ($P_3\geq 0.5$). We conclude that the redshift ambiguity of the BBGs is mainly driven by the existing depth at the 8.0$\mu$m band and that deeper data will be necessary to place more stringent constraints on their redshift distribution.

Finally, we note that several studies reported a significant fraction of MIPS 24$\mu$m detections among massive quiescent galaxies \citep{mancini09, marchesini10, nayyeri14, marsan15}. At $z$=3.0--4.5, the 24~$\mu$m samples $\lambda_{\rm rest}$$\approx$4--6~$\mu$m, where warm-hot dust continua or polycyclic aromatic hydrocarbons excited by star formation or AGN activity could contribute significantly to the flux. Exploration of such possibilities offers a promising avenue to learn about how the `red-and-dead' galaxies form and what roles AGN activity and nuclear starburst plays in the process. We notice the submillimeter (ALMA+SCUBA2) detection a fraction of an arcsecond away from a confirmed post-starburst galaxy has been reported recently \citep{glazebrook17}. However, given the shallow MIPS coverage in the Bo\"otes field (5$\sigma$ detection limit is 250$\mu$Jy), we are unable to quantify what fraction of our BBG candidates may harbor hidden AGN or starbursts. %The ambiguity of their physical nature calls for future JWST observations which can detect the presence of Balmer lines as direct evidence for their quiescence and probe the possibility of a hidden AGN or starburst.

\subsection{UV-bright Balmer Break Galaxy Candidates: Post-starburst or normal star-forming galaxy?}\label{psb}

The relatively strong Lyman break present in the seven optically bright BBGs places their redshift in the range $z_{\rm phot}$=3.6--4.0, giving us confidence that the $H$ and $K_S$ bands  straddle the Balmer/4000\AA\ break. The overall chi-square distribution obtained from our SED fitting procedure suggests that either delayed or exponentially declining SFH models with relatively short $\tau$ values (100--300~Myr) are preferred over constant SFH models, where the latter typically returns larger $\chi_r^2$ values. The median fit value are $\log{[M_{\rm star}/M_\odot]}$=11.0~($\sigma$=0.2) in stellar mass, 145~$(\sigma=42) ~M_\odot {\rm yr}^{-1}$ in SFR, and 433~Myr ($\sigma=$23~Myr) in population age.  In comparison, the CSF model returns higher values of SFR  205~$(\sigma=67)~M_\odot {\rm yr}^{-1}$ but similar stellar masses and ages. These values are also consistent with the stacking results shown in Table~1.

In Figure~\ref{stacked_sed} (right), we show the stacked photometry together with the best-fit SEDs assuming CSF  (blue) and exponentially declining (red) models. The overall SED shapes are very similar in the entire range of the rest-frame UV-to-IR wavelengths with the exception of the $K_S$ band sampling the rest-frame 4500\AA. A zoom-in on the wavelength range near the Balmer/4000\AA\ break is shown in the far right panel. %Thus, they may be simply more dust reddened star-forming galaxies. 

We also consider a scenario in which the galaxies are composed of two stellar populations formed at different times where the old population dominates the rest-frame optical emission while the UV emission originates from newly formed stars \citep[e.g.,][]{kriek09}. We explore a range of `double burst' models as follows: the SFHs of both populations are modeled as exponentially declining functions with  $\tau$ values ranging in $\tau=10-1000$~Myr. The ages of the two populations are also allowed to vary. The fraction of stellar mass formed in the second burst relative to the old population, $f_{\rm new}$, is varied from 0.01 to 0.50. The minimum $\chi^2$ is achieved at $f_{\rm new}\approx 0.05$ where a 200~Myr-old new burst is currently forming stars at rates of 114~$M_\odot {\rm yr}^{-1}$ (green line in Fig~\ref{stacked_sed}, right panel). The  $\chi^2$ values are similar out to  $f_{\rm new}\lesssim 0.1$, but increase more rapidly at $f_{\rm new} \geq 0.2$ ($\Delta \chi^2 = 0.4$ and 0.9 at $f_{\rm new}=0.2$ and 0.3, respectively). Thus, we conclude that the mass formed during the recent SF episode is small  ($<$10\%) compared to that of the evolved stellar population. The stellar population parameters obtained for all three different star formation histories are listed in Table~1.

Finally, we consider the possibility that the seven optically bright candidates are normal star-forming galaxies misclassified as BBGs due to the contamination of the $K_S$ band flux by  strong nebular emission such as [O~{\sc iii}] and H$\alpha$ \citep{shim11,stark13,schenker13}. Of particular interest to the present sample is  the [O~{\sc iii}]~$\lambda\lambda4959,5007$ doublet, which falls into the $K_S$ band at $z=3.1-3.6$. 
For all but two, the  redshift PDF peaks at $z\geq 3.7$ even though the majority has a non-zero probability of lying in the range $z_{\rm phot}=3.5-3.6$. For the remaining two, the peak of the PDF is $z\sim3.6$. \citet{schenker13} measured the  rest-frame [O~{\sc iii}] equivalent widths (EWs), and determined the median value of 280\AA\ \citep[see also,][]{holden16}. At $z=3.5$, it would lead to a substantial overestimation of the $K_S$ band continuum flux by 0.37~mag. 

However,  \citet{malkan17} recently noted that a strong anti-correlation exists between [O~{\sc iii}] EW and stellar mass such that the EW could be as low as 80\AA\  in $M_{\rm star} \approx 10^{10}{\rm M}_\odot$ galaxies. The median stellar mass of our UV-bright BBGs is nearly an order of magnitude larger than this value.  Similarly, only two galaxies in the Schenker et al. sample have UV brightness similar to our sample\footnote{The $z_{850}$ magnitudes of the Schenker et al. galaxies are 24.4 and 25.3~AB; the magnitude range of our UV-bright BBGs is $I=25.2\pm0.4$ and $Y=25.0\pm0.6$~AB.} whose EWs are 100\AA\ and 150\AA\ corresponding to a much less severe contamination of $\Delta m$ of 0.15 and 0.21~mag, respectively. The trend of decreasing EWs with increasing mass and  UV luminosity is likely  the same, given the relatively tight correlation between the two quantities among star-forming populations \citep[e.g.,][]{stark09,lee12a,salmon15}. 

We measure the mean [3.6]$-$[4.5] color to be $0.24\pm0.16$; \citet{stark13} reported  that the median [3.6]$-$[4.5] color is $\approx -0.23$~mag for $3.8<z<5.0$ galaxies, significantly bluer than the $\sim0.1$~mag color measured in their sample for the systems at $3.1<z<3.6$.  The color difference is attributed to the presence of strong H$\alpha$ emission in the former. The lack of excess 3.6$\mu$m band flux corroborates the possibility that nebular line contamination is not significant.

Given the color degeneracies between the above possibilities, discriminating a young post-starburst  from a rejuvenated old galaxy will be harder, requiring detection of their respective spectroscopic signatures; these will include Balmer absorption lines for post-starburst galaxies \citep[e.g.,][]{kriek09, glazebrook17} and nebular lines such as [O~{\sc ii}], [O~{\sc iii}], H$\beta$, and H$\alpha$, from the H~{\sc ii} regions. Future James Webb Space Telescope spectroscopy will help resolve this issue unambiguously \citep{Kalirai18}.

Regardless of their nature, we have uncovered a rare population of ultra-massive galaxies ($\gtrsim 10^{11}M_\odot$) which may have recently halted their star formation, or are nearing the end of their star-formation activity. 

A summary of all the 43 BBG candidates is given in Table~2.

\section{A Massive Galaxy Overdensity?}\label{sky_distribution}

\begin{figure*}[ht!]
\epsscale{1.2}
\plotone{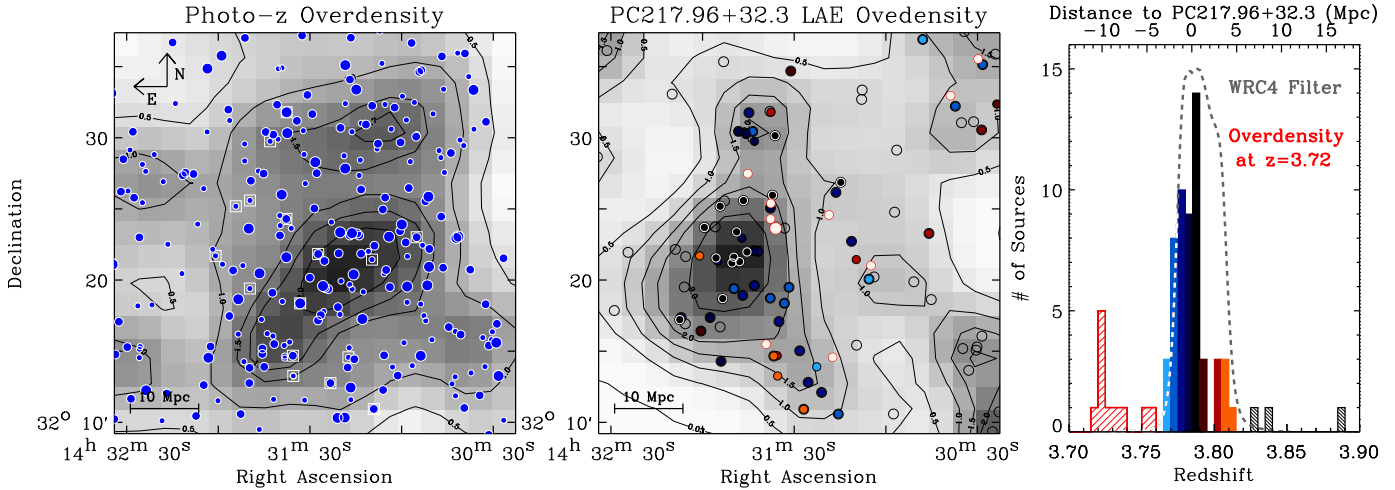}
\caption{
{\it Left:} Grey shades and density contours show the distributions of photo-z  member candidates (blue filled circles). White boxes indicate sources with known spectroscopic redshifts. The surface density map is created by smoothing the positions of each galaxy by a 4.7\arcmin-FWHM Gaussian kernel.  Symbol sizes indicate galaxy's stellar masses as $M_\star < 10^{10.5} {\rm M}_\sun$ (small), $10^{10.5} {\rm M}_\sun < M_\star < 10^{11} {\rm M}_\sun$ (medium), and $M_\star > 10^{11} {\rm M}_\sun$ (large). A comoving distance scale is indicated on bottom left corner.
{\it Middle:} density contours show the LAE distributions.  The spectroscopic members are color-coded by redshift indicated in the right panel. Six LAEs near the lower redshift cutoff of our LAE selection ($z=3.770-3.804$) lie close to the photo-$z$ overdensity peak.    
{\it Right:} Histogram of spectroscopic sources at $z=3.70-3.90$. Top abscissa indicates the corresponding line-of-sight distance (physical) measured from the structure redshift at $z=3.783$. The LAE redshift selection function (dashed line) is converted from the narrow-band filter bandpass. A smaller overdensity at $z\approx 3.72$ (red hatched histogram) is identified from our spectroscopic survey; the locations of these sources are indicated in the middle panel as white symbols outlined by red circles. Among them is G6025 -- an unusually large (20~kpc) galaxy at $z=3.72$ reported by \citet{lee18} -- shown as the largest circle. The three sources  at $z>3.82$ (dark hatched histogram) are not LAEs and thus are not used in our analysis. 
}\label{fig7}
\end{figure*}

\subsection{Sky Distribution of Protocluster Candidates} 
We show the sky distribution of the photo-$z$ protocluster candidates in the left panel of Figure~\ref{fig7}; the surface density enhancement relative to the mean density is shown as both greyscale and contour lines. The true density enhancement is expected to be higher as the mean density computed from the entire field includes the galaxies in the overdense region. 
 There is a clear indication of a large overdensity slightly south of the field center, outlined by the $1.5\bar{\Sigma}$ and $1.7\bar{\Sigma}$ lines. A smaller less significant one is found north of the field center. 
%The true density enhancement should be larger than 50\% relative to the mean density because the latter is computed from the entire field, which is dominated by the galaxies in the overdense region. 

In the same figure,  the sky distribution of known members of PC217.96+32.3 is shown in the middle panel; spectroscopic sources (which include both LAEs and LBGs) are color-coded by redshift. The density contour of the protocluster is constructed as before, but only using the LAE positions. Because our spectroscopic efforts were heavily focused on the LAE overdensity region, including the non-LAE members in the density calculation would artificially increase the overdensity. 

Comparing the density maps of the photo-$z$ and of protocluster LAEs, it is evident that they are not co-spatial. We perform a two-dimensional Kolmogorov-Smirnov (K-S) test \citep{peacock83,fasano87} to compare the the photo-$z$ distribution with the LAE distribution, and find the $p$-value of $2.9\times10^{-7}$. Thus, it is extremely unlikely that they are drawn from the same parent distribution at random. The 2D K-S test has also been used in \cite{Kuiper12} to compare between different structures. 

The largest photo-$z$ overdensity runs in the NW-SE direction. While it partially overlaps with the southern end of PC217.96+32.3, it stretches further west to the region devoid of the LAEs. A smaller and less significant overdensity lies just north of the main overdensity, which also overlaps slightly with a small LAE group north of the main LAE overdensity.  The larger overdensity is also closer to PC217.96+32.3, and thus most likely to have a physical connection to the confirmed protocluster. Being separated from each other,  the physical association of the two photo-$z$ overdensities is unclear. Thus, we focus on the larger overdensity in this work. 

The shapes and locations of the photo-$z$ and LAE surface overdensities are suggestive of a possible physical connection. One hypothesis is that  they are part of a single structure where the photo-$z$ overdensity lies in the foreground of the LAEs (i.e., at $z<3.76$), and as a result any Ly$\alpha$ emission from galaxies in this region is missed by the LAE selection filter. In the right panel of Figure~\ref{fig7}, we show the narrow-band filter bandpass converted to the redshift selection function (dashed line) together with the distribution of all spectroscopic sources in the range of $z=3.70-3.90$. Most of the  known members residing in the LAE overdensity lie at $z=3.775-3.785$, i.e., the blue half of the filter response. The southern end of PC217.96+32.3 is composed of galaxies in the redshift range where the filter response falls off steeply. Existing spectroscopy reveals that three LAEs there have the line centroids outside the narrow-band filter, but are selected as LAEs because of their high line luminosities and broad line widths. The high concentration of $z\approx 3.77$ LAEs where the two overdensities overlap provides a circumstantial evidence that the LAEs only partially trace the true extent of a single very large structure.

Another possibility is that the photo-$z$ overdensity is located further in the foreground of PC217.96+32.3 near an LBG overdensity at $z=3.72$. Of the ten galaxies at $z=3.721\pm0.04$ within our NEWFIRM coverage, three  reside within the $\Sigma=1.7\bar{\Sigma}$ region, and additional five lie just outside the $\Sigma=1.5\bar{\Sigma}$ contour line. The significance of this spectroscopic overdensity is difficult to assess given the limited extent and depth of the existing spectroscopy.
%, which has been primarily focused on confirming redshifts of the LAE candidates. As discussed in \citet{dey16}, photometric LBGs were given secondary priorities on multi-object spectroscopy performed with Keck/DEIMOS with an average integration of just 1.5~hr. As a result, 
All confirmed LBGs --  including the eight galaxies at $z\approx 3.72$ -- either have relatively strong Ly$\alpha$ emission or high UV continuum luminosities.  Further lending support to this possibility is G6025 (large white circle in Figure~\ref{fig7}), one of the eight galaxies that is unusually large   \citep[end-to-end length of $\sim$20~kpc:][]{lee18}. The ground-based morphology and large angular size are consistent with two UV-luminous galaxies involved in a major merger, a type of event that should occur more frequently in a dense environment.  

Finally, the photo-$z$ overdensity may represent a protocluster with no physical connection to PC217.96+32.3. The stacked redshift probability density function of protocluster candidates peaks at $z\approx 3.75$, but its width is not narrow enough to rule out the possibility that the true peak may be $\Delta z \gtrsim 0.1$ away from the peak value, which would place the structure at $\gtrsim 20$~Mpc away from PC217.96+32.3. 

Multiple protoclusters in close proximity are unlikely, but not impossible. \cite{Kuiper12} noted that there may be two separate galaxy overdensities near MRC~0317-257, a radio galaxy at z=3.13.  Similarly, a string of galaxy overdensities in the COSMOS field was found spanning over a line-of-sight distance of $\sim25$~Mpc \citep[$z=2.42$, 2.44, 2.47, and 2.51 reported by][respectively]{diener15, chiang15,casey16, wang16}. With the limited  spectroscopy, their physical connection  remains unclear. Two additional LAE overdensities of smaller magnitudes exist just outside our NEWFIRM field north of PC217.96+32.3 \citep{lee14},
%\citet{lee14} reported the existence of two additional LAE overdensities of smaller magnitudes, $\approx 8-10$~Mpc (physical) from PC217.96+32.3 in the transverse direction.
one of which was  confirmed to be a galaxy overdensity \citep{dey16}. 
%Both are located outside our NEWFIRM field, north of PC217.96+32.3. 
In \S~\ref{mil}, we return to this topic and evaluate the likelihood of multiple protoclusters in our survey volume.

\subsection{Sky Distribution of Balmer Break Galaxy Candidates}\label{bbg_distribution}

\begin{figure*}[ht!]
\epsscale{0.86}
\plotone{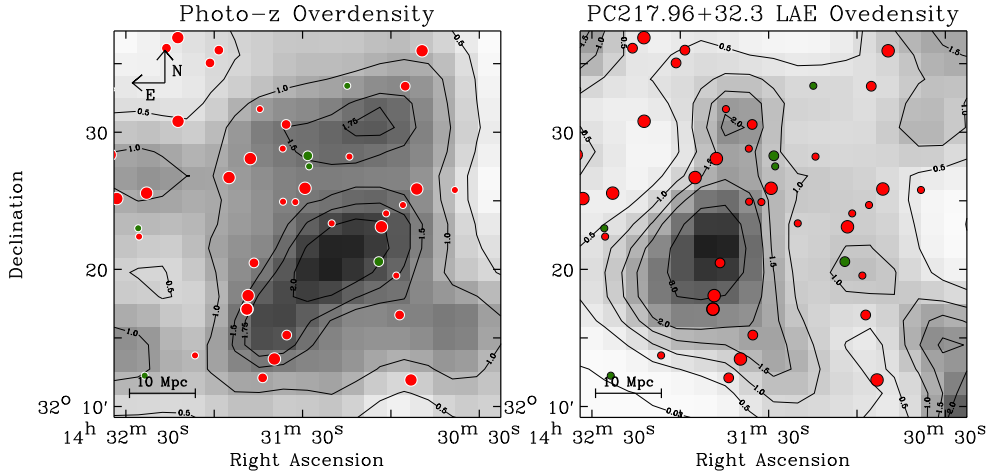}
\caption{
The sky positions of the BBG candidates are overlaid on the density contours of photo-$z$ (left) and of the LAE members of PC217.96+32.3  (right). Galaxies whose SEDs are consistent with young post-starbursts (see \S~\ref{sec_bbg}) are shown in green, and quiescent galaxy candidates are indicated in red. None of the quiescent galaxy candidates is included in our photo-$z$ sample. Large, medium, and small symbol sizes denote their estimated stellar masses corresponding to $\geq 2.5\times 10^{11}M_\odot$, $(1.2-2.5)\times 10^{11}M_\odot$, and $<1.2\times 10^{11}M_\odot$, respectively.
}
\label{density_BBG}
\end{figure*}

%In Figure~\ref{density_BBG} we show the locations of the 44 BBG candidates overlaid on the %photo-$z$ (left) and LAE (right) density maps. The distribution of the protocluster LAEs and %that of the BBGs are clearly dissimilar. The latter appears to almost systematically avoid %the LAE-rich regions: only one quiescent galaxy candidate lies near the protocluster core, %and two additional sources are near the $3\bar{\Sigma}_{\rm LAE}$ line. In contrast, the %photo-$z$ and BBG candidates generally trace one another reasonably well. We perform a 2D %K-S test using the BBG and LAE distributions, and find the $p$-value of 0.0002, while the %same test using the BBG and photo-$z$ distributions result in $p=0.07$, which confirms our %visual impression.

Several previous studies have reported a high concentration of galaxies dominated by old stellar populations near known massive protoclusters \citep{steidel05,kubo13,kubo15,wang16}. While those galaxies still await spectroscopic confirmation, such information would have important implications to the formation histories of massive cluster ellipticals. In this context,
we investigate whether our BBGs are physically associated with the structure revealed by the photo-$z$ overdensity. %Given the above tests and consider only a few BBGs locate in the photo-$z$ overdensity, this question remained open and call for the need of future spectroscopic observations. 

In Figure~\ref{density_BBG} we show the locations of the 43 BBG candidates overlaid on the photo-$z$ (left) and LAE (right) density maps. The BBGs seem to avoid the most overdense regions of both the LAEs and photo-$z$ candidates. Only one quiescent galaxy candidate lies near the LAE core, and two additional sources are near the $3\bar{\Sigma}_{\rm LAE}$ line. Only three BBG candidates locate near the $2\bar{\Sigma}_{\rm photoz}$ contour line. The relative void of all types of galaxies potentially associated with the structure in the southern corner and northern end of the field is also noteworthy. The fact that the same regions are well populated with lower-redshift sources perhaps suggests that the void is not artificially created by  the presence of bright  sources such as saturated stars or large galaxies.

We perform a 2D K-S test using the BBG and LAE distributions, and find the $p$-value of $3\times10^{-4}$,  indicating the significant disparity between the spatial distributions of the two samples. The same test using the BBG and photo-$z$ distributions result in the $p$ value of $0.05$.
The K-S test evaluates the similarity of two univariate samples by constructing their cumulative distributions and computing their maximum distance. Because multivariate samples can be ordered in more than one way, multi-dimensional K-S tests lack the statistical rigor of the 1D test, and thus need to be understood in the context of carefully controlled tests. To this end, we populate the survey field with two {\it a priori} known distributions and perform the 2D K-S test to quantify the range of the $p$ values.  For each test, we create 1,000 separate realizations. 

First, we create two random distributions, each matching the number of BBGs and photo-$z$ sources. We obtain the mean (median) $p$ value of 0.30 (0.27) with the standard deviation ($\sigma$) of 0.21. Second, we compare the photo-$z$ sample with a randomly chosen subset of itself consisting of 43 sources (matching the number of BBGs), which result in the $p$ value of 0.37 (0.36; $\sigma=0.21$). These tests suggests that we can reliably rule out the possibility that the LAEs and BBGs -- having $3\times10^{-4}$ --  are drawn from the same parent distribution. 

The relationship between the BBGs and photo-z candidates, however,  is less clear with $p=0.05$. Comparison of the photo-$z$ distribution with 43 randomly distributed  sources yields the $p$ value of 0.11 (0.07, $\sigma=0.12$), comfortably bracketing the measured $p$ value. Thus, we cannot statistically rule out the possibility that the BBG positions are not correlated with those of photo-$z$ members. Spectroscopic redshifts are necessary for progress.

\subsection{Estimate of true overdensity and descendant mass}\label{overdensity}

We assess the significance of the structure by estimating the range of the true galaxy overdensity given the observed level of the surface density enhancement. The transverse size of the photo-$z$ overdensity is computed by interpolating the 1.5$\bar{\Sigma}$ iso-density contour, which yields 139~arcmin$^2$ or 26.4~Mpc$^2$ (physical) at $z=3.78$. Since  physical scale remains constant within 2\% at $z=3.65 - 3.85$, our subsequent estimate of the overdensity and masses should be relatively insensitive to the  precise redshift of the structure. We assume that the line-of-sight distance from the front to back of the structure is 15~Mpc; this is motivated by the fact that the effective diameter of the progenitors of massive present-day clusters lies in this range \citep{chiang13}. The redshift distribution of the known members of PC217.96+32.3 ranges over $z=3.77-3.79$ is  consistent with this expectation (see the right panel of Figure~\ref{fig7}).

We infer the range of the intrinsic galaxy overdensity by performing Monte Carlo simulations as follows. In each run, we create a mock field containing one protocluster with a galaxy overdensity $\delta_g$ in the middle by populating points randomly in the $(\alpha,\delta,z)$ space. The ``protocluster region'' is defined as a rectangle. The overall number of sources and the transverse area of the protocluster match those of the data.  An intrinsic galaxy overdensity, $\delta_g$,  is  chosen at random in the range $\delta_g=1-20$. We divide the redshift range $[3.4,4.2]$ into 40 bins with a binsize of $\Delta z=0.02$, corresponding to the stepsize of 15~Mpc in comoving line-of-sight distance. Taking $\delta_g$ as  intrinsic overdensity, the number of true members is $N_{\rm proto}=(1+\delta_g)N_{\rm phot}/(40+\delta_g) $ where $N_{\rm phot}$ is the total number of observed protocluster candidates in the field, and populate them at random within the protocluster region. Setting $\delta_g=10$ (5) means that 58 (35) galaxies are part of the structure. The remainder  ($N_{\rm phot}-N_{\rm proto}$) are assigned randomly assuming a uniform distribution in both transverse and line-of-sight positions. We  construct the surface density map of the mock image using the identical procedure as described previously, and  estimate the mean surface density enhancement within the protocluster region. 

We repeat the above procedure 10,000 times and obtain the empirical  relation between the true overdensity and surface overdensity. The scaling relation is well-behaved and nearly linear. Given the observed surface overdensity (the mean enhancement is 1.81 within the 1.5$\bar{\Sigma}$ iso-density contour), we estimate that the intrinsic overdensity of the structure is $\delta_g=5.5-10.2$ with the median value of 7.8. The value is comparable to the redshift overdensities found for several known protoclusters. \citet{steidel05} measured a redshift overdensity of $\delta_g\sim7$ for a $z=2.30$ structure. Based on the VIMOS Ultra Deep Survey \citep{lefevre15},  \citet{lemaux14} and \citet{cucciati14}  reported the inferred redshift overdensity of $\delta_g=10.5\pm2.8$ and $\delta_g=12\pm2$  for a protocluster at $z=3.28$ and $z=2.90$, respectively.  These values are larger than that determined for the SSA22  protocluster at $z=3.09$, $\delta_g\sim3.5-4.0$ \citep{steidel98,steidel00,Hayashino04,matsuda05,Yamada12,Topping18}. 

To investigate how sensitive the inferred $\delta_g$ value is to the transverse area of the surface overdensity, we rerun the simulations using the iso-density of 1.3$\bar{\Sigma}$ and 1.6$\bar{\Sigma}$; lowering the density contrast effectively increases the effective area, while raising it has the opposite effect. We find that the estimate of the underlying overdensity is relatively insensitive to a specific choice of density contrast used to estimate the extent of the structure. However, our test does not include the possibility that the surface overdensity is systematically overestimated  either due to Poisson shot noise or a superposition of another unrelated group of galaxies. Given the lack of spectroscopy in the region, we currently have no way of quantifying this possibility. If the surface overdensity region is reduced by 20\%, the $\delta_g$ value would decrease to 3.9--7.0. 

Based on the inferred galaxy overdensity $\delta_g$, we estimate the descendant mass of the underlying structure, i.e., the total mass enclosed within the overdense region which will be gravitationally bound and virialized by $z=0$, which can be expressed as:
\begin{equation*}
\label{mass_estimate}
M_{z=0}=(1+\delta_m)\langle\rho\rangle V
\end{equation*}
where $\langle \rho \rangle$ is the average matter density of the universe ($=[3H_0^2/8\pi G]\Omega_0$),  $\delta_m$ is the  matter overdensity, and $V$ is the comoving volume of the galaxy overdensity. With the adopted cosmology, Equation \ref{mass_estimate} is equivalent to $M_{z=0}=[3.67\times 10^{10}M_\sun]~ (1+\delta_m)~ [V/(1~{\rm cMpc})^3]$. The two overdensity parameters, $\delta_g$ and $\delta_m$, are related through the equation $1+b\delta_m = C(1+\delta_{g})$ where   $C$ denotes a factor correcting for the effect of redshift-space distortions \citep{steidel98}, and $b$ is galaxy bias. Given the lack of details to assume otherwise, we use the $C$ in the case of spherical collapse: $C(\delta_m,z)=1+\Omega_m^{4/7}(z)[1-(1+\delta_m)^{1/3}]$. As for galaxy bias, we adopt $b=3.5$. Our choice is justified by the fact that the majority of our photo-$z$ sources lie in the observed UV luminosity range comparable to those of $L\gtrsim L^*_{\rm UV}$ LBGs at $z=3-4$. The bias value of the latter has been estimated through measurements of their clustering properties \citep[e.g.,][]{ouchi04b, lee06,hildebrandt09}. We solve the above equations for $\delta_g$ and use Equation (1) to obtain the mass estimate. 

The enclosed mass in the photo-$z$ structure is $(7.9\pm1.0)\times 10^{14}M_\odot$ given the overdensity $\delta_g$ of $7.8\pm2.4$. The inferred dark matter overdensity is $\delta_m=1.39\pm0.3$.  Increasing the bias value to $b=4$ would decrease the mass by 10\%. 

\section{Discussion}

\subsection{The prevalence of massive quiescent galaxies in protocluster environment}\label{bbg_discussion}

% a paragraph estimating the number density of massive  galaxies in the field; 
% - make a simple estimate above some mass threshold.
% - compare the number density with the SMF estimates from the field. 
% - compare it with Kubo et al. 
% - compare the results with other papers, including Lemaux+(2014), Steidel+ (2005), etc. etc. 

We evaluate how the  number of massive quiescent galaxies ($\geq 10^{11}M_\odot$) in our field compares with that expected in an average field. Based on $K_S$-selected galaxies in the 1.6~deg$^2$ COSMOS/UltraVISTA field, \citet{muzzin13} estimated that at $z=3-4$,  the cumulative number density of galaxies with  $M_{\rm star}\geq 10^{11}M_\odot$ is $(1.4^{+2.2}_{-0.5})\times 10^{-6}$~Mpc$^{-3}$. In our survey field (28\arcmin$\times$35\arcmin), one expects to find $2.5^{+3.9}_{-0.8}$ BBG-selected quiescent galaxies. Similarly, \citet{spitler14} identified 6 quiescent galaxies above $M\geq 10^{11}M_\odot$ in the ZFOURGE survey corresponding to the surface density of $0.015\pm0.006$~arcmin$^{-2}$, such that $3.7\pm1.5$ quiescent galaxies are expected in our field. We assume in the above calculations that the selection function takes the form of a top hat filter in the range $z=3.6-4.2$ where the $H-K_S$ color samples the Balmer/4000\AA\ break. The relative change of angular diameter distance in this range is 6\%, and should result in 12\% in the expected number depending on the redshift distribution of BBGs. 

Taking the \citet{muzzin13} measurement as the field average, the implied overdensity of massive quiescent galaxies is $\delta\Sigma_{\rm BBG} \sim 16$! Excluding all of our post-starburst BBG candidates (assuming all are strong [O~{\sc iii}] emitters at $z\sim3.4$), the remaining BBGs correspond to $\delta\Sigma_{\rm BBG}\approx 13$. Using the \citet{spitler14} estimates, the overdensity is $\delta\Sigma_{\rm BBG}=11$ (9) with (without) the potential [O~{\sc iii}] emitters.

We also compare the observed abundance of quiescent galaxies with that measured in the SSA22 protocluster at $z=3.09$. \citet{kubo13} used color criteria tuned to $z\sim3$ ($i^\prime - K>3$, $K-[4.5]<0.5$, and $K<23$), and identified 11 massive galaxies ($\gtrsim 10^{11}M_\odot$) concentrated near the overdensities of other types of galaxies with the surface density of $0.10\pm0.03$~arcmin$^{-2}$. In comparison, the overall surface density of BBGs in our field is $0.06\pm0.01$~arcmin$^{-2}$. Within a smaller rectangular region (15\arcmin$\times$16\arcmin) in which the surface density of photo-$z$ sources is enhanced by 50\% (Figure~\ref{density_BBG}, left), we find 21 quiescent BBGs there in, corresponding to the surface density of $0.09\pm0.02$~arcmin$^{-2}$. All errors are given assuming Poisson shot noise. Considering the change of angular diameter distance, the surface density per unit comoving transverse area  is $0.027 \pm 0.008$~Mpc$^{-2}$ and $0.021\pm0.004$~Mpc$^{-2}$ for the SSA22 and the present structure, respectively.  
Similarly, \citet{lemaux14} estimated that the implied overdensity of massive ($\geq 10^{10.8}M_\odot$) red galaxies in a $z=3.29$ protocluster is $\delta_g=25.1\pm15.2$.  

A large population of massive quiescent galaxies found in our field implies that the formation of cluster galaxies occurred in shorter timescales and at earlier times than the field galaxies. Our results confirm an early onset of cluster red sequence \citep[e.g.,][]{kodama07,lemaux14}. This is in a broad agreement with star formation histories of present-day cluster ellipticals inferred from absorption line studies \citep{thomas05}. Little to no evolution of the cluster red sequence out to $z\sim 1.4$ further strengthens this expectation \citep[e.g.,][]{blakeslee03,mei06}. 

\begin{figure*}[t]
\epsscale{1.1}
\plotone{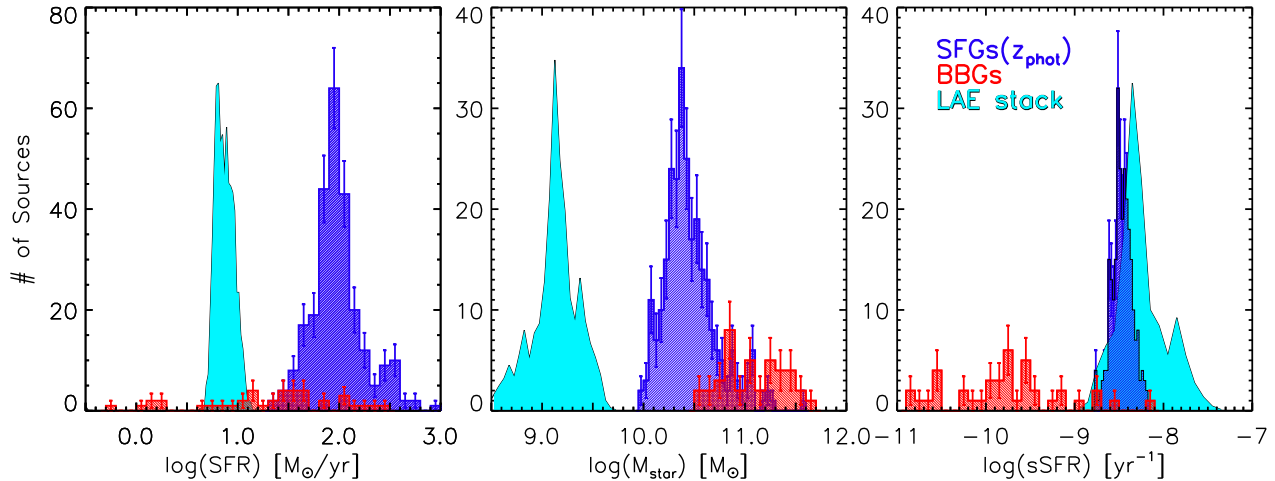}
\caption{
The distributions of star formation rates (left), stellar masses (middle), and specific SFRs (right) are shown for the LAE protocluster members (cyan) and  photo-$z$ protocluster candidates (blue) and BBG candidates (red). The errors reflect the Poisson uncertainties. For clarity, finer binsizes are used for the photo-$z$ sources than for the BBGs. As for the LAEs,  stellar population parameters are derived from bootstrap realizations of image stacking analyses (see text).  In all panels, the LAE distribution is rescaled to have the same peak height as the photo-$z$ members. 
%Together, these galaxies span several orders of magnitude in all probed quantities. 
}
\label{sfr_mass}
\end{figure*}

The sky distribution of BBGs appears to trace the full extent of the large scale structure rather than being concentrated in the highest density environments. Few are found in either LAE or photo-$z$ overdensity peaks (see \S\ref{bbg_distribution}). We speculate that BBGs may be the central (and most massive) inhabitants of the massive halos that are in the process of merging. The implication is that they were quenched long before the final coalescence of the structure which occurred much later. Therefore, the quenching of massive cluster ellipticals is caused by the early onset of the `mass quenching' rather than by any environmental effect suppressing their formation \citep{peng10}. This is in line with a study of intermediate-redshift galaxy clusters by \citet{brodwin13}, who found that the level of star formation in cluster environment declines below that in the average field only at $z\lesssim 1.4$ \citep[also see,][]{tran10}. Recent discoveries of compact galaxy groups in protocluster environments support  this view, as a fraction of quiescent galaxies in such a group is observed to be low \citep[e.g.,][]{wang16,kubo16}.

\subsection{Diverse types of galaxies tracing a massive protocluster}

In this work, we have identified protocluster member candidates by employing two selection methods, namely photo-$z$ selected star-forming galaxies and Balmer break galaxies. When combined with a population of LAEs   in the same field \citep{lee14, dey16}, these  samples showcase  diverse types of inhabitants residing in a very overdense cosmic structure.

In Figure~\ref{sfr_mass}, we show SFR, stellar mass, and sSFR values measured for our sample galaxies. The estimates for the photo-$z$ candidates are made on individual galaxies. As for the quiescent BBG candidates, we fix the redshift to $z=3.8$ for the SED fitting (see \S~\ref{bbg_q} for discussion on redshift degeneracy); for the UV-bright BBGs with robust photo-$z$ estimates, we fix the redshift to the best-fit value.

As for the LAEs, while they have robust redshift estimates, they are too faint at infrared wavelengths to yield robust estimates of stellar population parameters on an individual basis. Instead, we perform image stacking on their positions, and measure the parameters based on the aperture photometry on the stacked images. A total of  150 LAEs are used for stacking analysis after removing those too close to nearby bright sources.
To estimate the range of their physical parameters, we randomly draw a subset of the LAEs, and perform image stacking, aperture photometry, and SED fitting procedure.  Their distributions of stellar population parameters shown in Figure~\ref{sfr_mass} are based on 2,500 such realizations. Since median stacking is insensitive to significant outliers, the distribution of their physical parameters should be taken as a lower limit rather than the full range spanned by the LAEs.

The sample galaxies span a wide range of SFRs and stellar masses: the lack of overlap is at least in part driven by the selection effect. The lack of photo-$z$ candidates at $M_{\rm star}\lesssim 10^{10}M_\odot$ is tied  to the sensitivity of our $K_S$ band data. A 10$\sigma$ detection ($K_{S,\rm{AB}}$=24.0) corresponds to the rest-frame optical luminosity of a $z=3.8$ galaxy with stellar mass $\approx 10^{10.2}M_\odot$, assuming an exponentially decaying star formation history with the $\tau$ value of 0.5~Gyr.  The paucity of galaxies with ${\rm SFR}\lesssim50~M_\odot {\rm yr}^{-1}$ is also driven by the same mass limit, given the correlation between SFR and $M_{\rm star}$. The large median mass of the BBGs is driven by the IRAC color selection  as discussed in Sec~\ref{selection_bbg}.  The steep decline in the number of galaxies at ${\rm SFR}\gtrsim 150~M_\odot {\rm yr}^{-1}$ \citep[e.g.,][]{smit12}  is likely further helped by the photo-$z$ selection which is biased against redder (dustier) galaxies than typical LBGs.
The intrinsic distribution of these parameters spanned by different types of galaxies remains uncertain: such information will require careful analyses of deeper multiwavelength data and the modeling of their respective selection biases, which are outside the scope of this paper.

The measured overdensities of different galaxy types highlight how they trace the same underlying large scale structure(s). Such measures are more robust against any selection biases mentioned previously as any such bias should apply equally to field and cluster galaxies, and thus should minimally impact their spatial distributions. The observed surface overdensity of photo-$z$ galaxies is  $\delta \Sigma_{\rm phot}\approx 1.5$, similar to that of the LAEs over the same general area. However, we show in \S~\ref{overdensity} that the spatial overdensity of the photo-$z$ galaxies is  much larger, $\delta_g=7.8\pm2.4$, than that of the LAEs. This is  because the former is distributed over a much larger line-of-sight distance (i.e., larger $\Delta z$), and as a result, its surface overdensity is substantially diluted by the interlopers. It is also possible that the narrow band Ly$\alpha$ filter `misses' the core of the protocluster, and is only picking up the outer parts of the protocluster. 
 In comparison, the surface overdensity of BBGs of the region is much higher at $\delta\Sigma_{\rm BBG}$$\approx$9--16. 
 
 If all types of galaxies we consider here (LAEs, BBGs, and photo-$z$ candidates) trace the same underlying structure represented by a matter overdensity $\delta$, the implication would be that more massive BBGs are far more biased tracers of the matter distribution than less massive star-forming galaxies. Our findings are consistent with the expectation from existing clustering studies, that more luminous/massive galaxies have larger biases \citep[e.g.][]{GD01,ouchi04b,adelberger05,lee06,gawiser07,guaita10,kusakabe18}. 

One corollary to the luminosity/mass-dependent bias is that,  everything being equal, low-mass low-bias galaxies such as LAEs are the least biased (thus most reliable) tracers of the  density distribution within the large-scale structure. Using LAEs to `map out' the protocluster environment has additional advantages including the relative ease of redshift identification through the narrow-band selection technique and the abundance of low-luminosity galaxies implied by the steep  faint-end slope of the UV luminosity function at this redshift range \citep[e.g.,][]{bouwens07,reddy09,alavi16,malkan17}. 
Given the difficulties of obtaining spectroscopic redshifts for faint distant galaxies, LAEs offer the best  practical means to survey the local environment of massive protoclusters, thereby allowing for studying the impact of local environment on its galaxy constituents \citep[e.g.,][]{kubo13,umehata14,kubo16}. 
While large numbers of protocluster candidates are being identified from wide-field deep surveys \citep[e.g.,][]{toshikawa16,toshikawa18}, the lack of narrow-band observations targeting these structures will remain a main challenge in utilizing these structures to elucidate the  physics in the main epoch of cluster formation. 
\begin{figure*}[t]
\epsscale{1.1}
\plotone{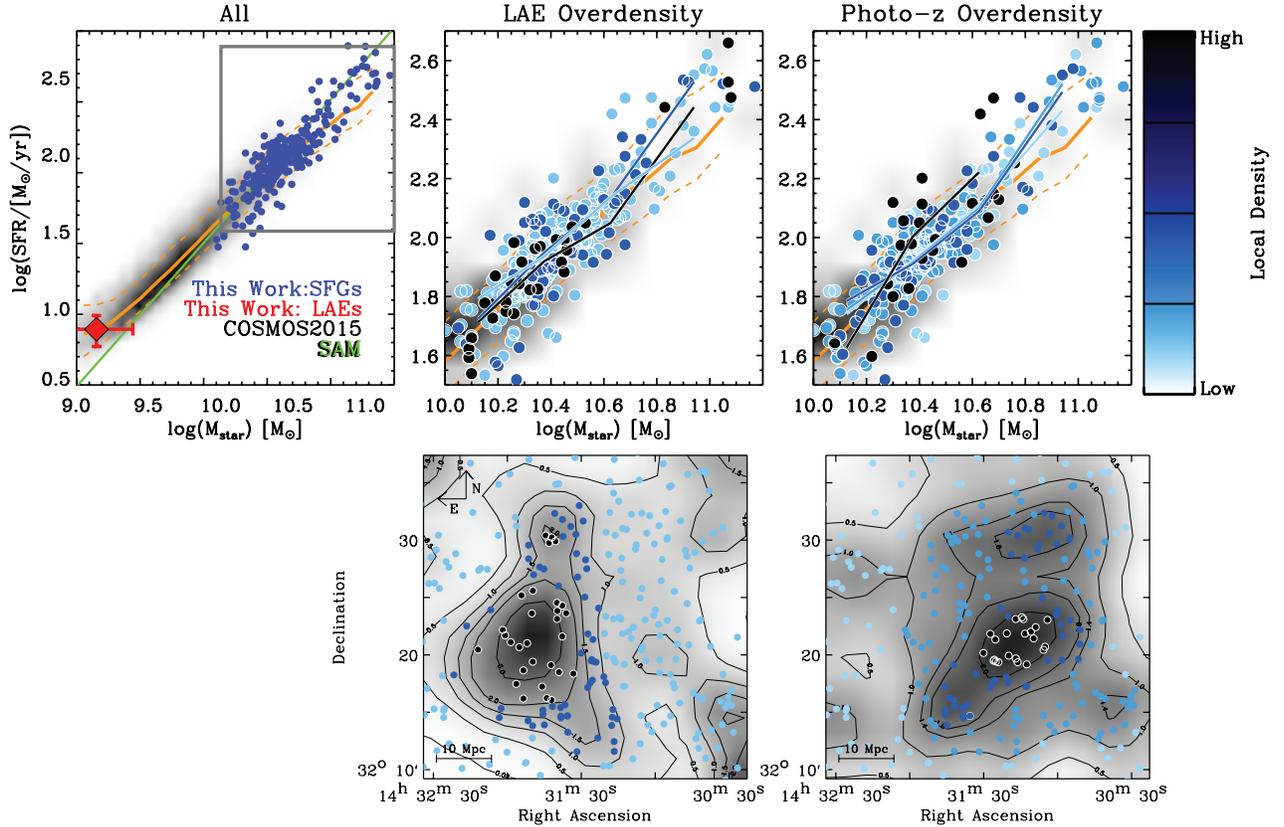}
\caption{
{\it Top left:} the location of photo-$z$ protocluster candidates (blue circles) and LAEs (red diamond) on the SFR-$M_{\rm star}$ space. The latter measurements are based on median image stacking. Grey scales show the distribution of COSMOS field galaxies at the same redshift range, and are color coded by the source density in each bin. 
The median scaling relation and the 16th/84th percentiles determined for the average field are shown in all top panels as solid and dashed orange lines, respectively. 
%
%The median scaling relation for the average field is determined from their distribution, and is shown as an solid orange line (the dashed orange lines denote the 1-$\sigma$ dispersion). 
A prediction from a semi-analytic model by \citet{dutton10} is indicated by a green line. 
{\it Middle and right panels:} A zoom-in on the parameter space populated by our photo-$z$ candidates. Each galaxy is color coded by its approximate local environment determined based on the surface density of the LAEs (top middle) and photo-$z$ member candidates (top right) such that darker shades represent higher environmental densities. The sky distribution of the same galaxies are shown in bottom panels with the density maps (identical to those in Figure~\ref{fig7}) overlaid. In both panels, the scaling relation is recomputed for each environmental density bin. The same scaling law is obeyed by all environmental bins although there is a hint that the galaxies residing in the highest photo-$z$ overdensities appear to have higher SFRs than the rest. 
}
\label{main_sequence}
\end{figure*}

\subsection{Impact of local environment on stellar populations}\label{env_impact}
% discuss the main sequence, compared to that observed in COSMOS.
% a paragraph about the difficulties in systematic investigation of environmental effects given few have confirmed redshifts, will be done in future works after spectroscopic and generally more data

The primary challenge in investigating the environmental effect on protocluster constituents is the lack of spectroscopic redshifts, which prevents unambiguous confirmation of cluster membership and inhibits a robust mapping of the density profile of the cluster. Because our selection methods target a relatively broad range of redshift, all galaxy samples are expected to contain and may be even dominated by interlopers not associated with the structure we wish to probe. 
%Given the redshift range our selection methods span, all galaxy samples --  with the exception of the LAEs -- are expected to contain and likely are even dominated by interlopers not associated with the structure we wish to probe. 
These considerations testify to the clear need of spectroscopic information in making progress.

One possible way to discern any environment trend is to compare the galaxy statistic measured in a protocluster field with that obtained in a field without any strong density enhancements.. Provided that the environmental effects are strong and a substantial number of galaxies in the sample  belong to the protocluster, a qualitative trend may be identified through this comparison \citep[e.g.,][]{cooke14}. 
However, a comparative study is only meaningful  if the two datasets are well matched in depth, dynamic range, and wavelength coverage, which determine the precision with which photometric redshifts and stellar population parameters of the galaxies can be measured.

With these caveats in mind, we compare the properties of protocluster candidates with those of a control sample. The control sample is constructed from the COSMOS15 catalog \citep{laigle16} where the sources whose best-fit photo-$z$ solution lies in the range $z_{\rm phot}=3.4-4.2$ are selected. After removing galaxies with multiple peaks in the photo-$z$ PDFs, the sample consists of 19,318 sources. We run the CIGALE software using the identical setup as previously,  assuming constant SFHs for the both samples. While it is unrealistic to expect that all galaxies have constant SFHs, we are interested in the comparison of the two samples and not in exploring the full behavior of galaxies. A different SFH choice  would generally shift measured quantities in the same direction for most galaxies, and thus would not change our conclusions. 
Finally, we note that the photo-$z$ precision for the COSMOS galaxies is expected to be much better ($\sigma/(1+z)\sim 0.02-0.03$) than for our sample ($\sigma/(1+z)\sim 0.06$) thanks to the better imaging depth and finer wavelength sampling in the optical/near-IR wavelengths. While the larger uncertainty can  introduce a larger scatter in the overall distribution of a derived quantity, it will not impact our ability to discern any mean relation between two different quantities.

%our slightly larger redshift uncertainty is unlikely to have a substantial impact on the overall  distributions of the derived parameters.  The expected cosmological dimming between $z=3.7$ and $z=4.0$ is only 0.13~mag. 

%While it is unrealistic to expect that all galaxies have constant SFH, our intention is to compare the two samples rather than to investigate 

%a different choice would generally shift measured quantities in the same direction for the majority of galaxies. 
% However, since we just want to compare the two samples, assuming the same constant SFH is enough for this purpose.

%Clearly, it is unrealistic to expect that all galaxies meet these assumptions. However, our simplistic assumptions are justified by the fact that we are mainly interested in the relative position of G6025 on this parameter space rather than exploring the full behavior of galaxies in general. Because the SFH of a galaxy cannot be well constrained, a different choice of the adopted SFH would not change the location of G6025 on the SFR?Mstar plane relative to the control sample

In the left panel of Figure~\ref{main_sequence}, we show the locations of our photo-$z$ sources  and of LAEs on the SFR-$M_{\rm star}$ plane together with those of the control sample.  A prediction from a semi-analytic model \citep{dutton10} is also shown. Both our photo-$z$ candidates and LAEs occupy the same region as the field galaxies, suggesting that they obey the same star formation `main sequence' scaling relation, consistent with existing studies \citep[e.g.,][]{koyama14,cucciati14,Erfanianfar16}. From the same figure, it is evident that the COSMOS datasets can probe galaxies down to much lower masses than the present dataset. The mismatch of the sensitivities of the two datasets renders it challenging to compare how the number counts in bins of SFR or stellar mass differ in the these samples. 

To investigate possible environmental trends, we divide the photo-$z$ sample into  several environmental bins and color-code them accordingly where darker shades represent higher densities.  Given the uncertainties in the extent and center of the structure, we define local environment using the LAE and photo-$z$ surface densities. The results are shown in top middle and right panels. The overall correlation --  measured for each subsample in mass bins of $\Delta\log{M_{\rm star}}=0.25$ -- is shown in solid lines. The SFR-$M_{\rm star}$ scaling laws measured from these subsamples are generally similar to that measured in the COSMOS sample.  

We detect a hint of enhanced star formation activity in the highest photo-$z$ overdensity subsample. Four galaxies deviate from the field average by 0.3-0.4~dex (a factor of 2--3). The overall scaling relation in this bin has a slightly higher normalization (i.e., $\sim$0.1~dex higher SFR in a given stellar mass bin) although the scatter is substantial. Interestingly, the same bin also lacks massive galaxies above $\log{M_{\rm star}}=10.8$. The high-mass high-SFR end is well populated by  galaxies residing in all environments. All in all, the environmental effects on star-forming galaxies appear to be minimal. 

%There is a hint of higher levels of SFR in the highest photo-$z$ overdensity subsample (black circles in the top right panel)  where four galaxies in the bin deviate from the mean relation by a factor of $2-3$ (0.3-0.4~dex). The median SFR in the mass range $\log{M_{\rm star}}$=10.3 --10.6 is slightly larger (0.1~dex) than the rest of the galaxies. The same bin also lacks galaxies with stellar masses larger than $\log{M_{\rm star}}=10.8$. 

%No clear environmental trend can be discerned. Unlike several existing studies \citep[e.g.,][]{cooke14}, we do not find an excess number of massive star-forming galaxies, a deficit of low-mass galaxies, or systematically higher sSFR in high-density environments. 
%The mean stellar mass and SFR values measured in each density bin are virtually identical at $\langle \log M_{\rm star}\rangle$ = $10.41\pm0.27$, $10.44\pm0.21$, $10.50\pm0.19$, and $10.46\pm0.22$ and $\langle \log {\rm SFR}\rangle$ = $1.96\pm0.26$, $2.02\pm0.20$, $2.05\pm0.26$, and $1.99\pm0.21$, from the highest- to lowest-density bins, respectively. 

The lack of detectable environmental effects on the galaxy properties is puzzling. Uncertain cluster membership surely plays a role in diluting any existing trend by misplacing a subset of galaxies into a wrong density bin. However, should there be an excess of high-mass or high-SFR galaxies in dense environments, our analyses would have captured it as the regions most likely to be dense are counted as such in one or the other scenario. Hence, our analysis suggests that the environmental effect on star formation is likely a subtle one. 

Alternatively, most of the enhanced star formation is perhaps obscured from our view by dust. 
\citet{koyama13} reported that while  sSFRs are higher for galaxies in cluster environment than those in the field,  the trend emerges only when the mid-infrared budget of the SFR is properly accounted for. They argued that their result may be explained if a higher fraction of nucleated dusty starbursts  exist in cluster environments where dust properties are significantly different from normal star-forming galaxies, such that applying the same dust correction as the field galaxies would underestimate the true SFRs.

The lack of extreme star-formers in our sample (in both field and protocluster) is also in part a selection effect. 
Extremely dusty starbursting galaxies would not be included in our sample, as they would not have a strong enough spectral break for us to identify them robustly, or perhaps, are entirely invisible in the optical or infrared wavelengths. We therefore cannot test for the prevalence of 
%Such a scenario would be in line with the prevalence of 
dusty starburst systems in dense environments reported by several studies \citep{hung16,casey16}. Testing these hypothesis will require deeper infrared and submillimeter coverage of the field. 

\subsection{Search for Extreme Sources in the Protocluster Field}
% look for extreme objects, X-ray, radio, etc. 
% explain the difficulties doing this analysis with the current data
% brightest cluster galaxy candidates?
% compare the masses of the most massive ones to those of local, and intermediate-z BCGs.

The presence of powerful radio galaxies has been used as a signpost of highly overdense regions \citep{venemans05, miley06, kajisawa06, venemans07, overzier08, hatch11a, Kuiper12,cooke14,rigby14}. Likewise, highly overdense structures appear to harbor powerful AGN observed as  X-ray or submillimeter luminous sources, or giant Ly$\alpha$ nebula \citep[e.g.,][]{steidel00,prescott08,lehmer09,hung16,casey15,casey16}. 

Motivated by these findings, we search the existing radio and X-ray source catalogs to look for a sign of enhanced AGN activity. We cross-match the {\it Chandra} X-ray point-source catalog of the Bo\"otes field \citep[XBo\"otes:][]{kenter05} with our photo-$z$, BBG, and LAE positions, and find no match.  The XBo\"otes survey sensitivity of the full band (0.5--8~keV) is $7.8\times10^{-15}$~ergs~cm$^{-2}$~s$^{-1}$. \citet{lehmer09} studied X-ray detected sources in and around the SSA structure at $z=3.09$, and found that the X-ray flux for the confirmed members range in $(0.3-5.0)\times 10^{-15}$~ergs~cm$^{-2}$~s$^{-1}$. Thus, non-detection merely suggests that even the brightest X-ray sources in SSA22 would lie below the XBo\"otes detection limit. 

%existing {\it Chandra} sensitivity in the Bo\"otes field is insufficient to detect typical X-ray sources. 

%Non-detection likely suggests that the existing sensitivity is insufficient. \citet{lehmer09} studied X-ray detected sources in and around the SSA22 structure, another protocluster at $z=3.09$; the X-ray flux for the confirmed members ranged in $(0.25-5.01)\times 10^{-15}$~ergs~cm$^{-2}$~s$^{-1}$. Given the higher redshift and survey sensitivity, none of these sources would have been detected in the XB\"otes catalog. 

We also search for radio counterparts of our protocluster candidates (both photo-$z$ and BBG candidates) in the radio source catalog based on deep Low Frequency Array (LOFAR) 150 MHz observations (Tasse et al. in prep). The rms noise of of the data is 59 $\mu$Jy/beam. Using the matching radius of 2\arcsec, no counterpart is found. We also compare our source positions against the photometric redshift catalog of the same LOFAR-detected sources constructed following the method presented in \citet{Duncan18b,Duncan18a}, which covers roughly two thirds of our survey field and only half of the photo-$z$ overdensity region. This is because the bottom one third of our survey field lies outside of the  NDWFS field \citep{ndwfs}. Once again, no credible counterpart is identified. In addition, we cross-match our candidates with deep Westerbork Synthesis Radio Telescope (WSRT) 1.4-GHz catalog covering Bo\"otes field \citep{Devries02}, and find no counterpart. Therefore, we can rule out the presence of any high-redshift radio source with the flux density $\gtrsim $ 0.2~mJy in the probed redshift range.

Apart from the limited survey sensitivities, non-detection of powerful AGN in the protocluster member candidates is perhaps not surprising. As discussed previously, the majority of our photo-$z$ candidates, by design, resemble LBGs with a clean spectral break. This requirement effectively removes all galaxies that are either dusty starbursts  or AGN with power-law-like SEDs similar to those identified by \citet{hung16} in the COSMOS field; robust identification of such galaxies will require improved sensitivities and wavelength baseline. %Using deeper spectroscopic observations scheduled in Spring 2018, we intend to further investigate the possible presence of AGN and radio galaxies in the field.  

\subsection{The plausibility of a very large structure}\label{mil}

We assess how likely it is to find both structures or one very large structure in our survey volume. We utilize a catalog containing 2,731 simulated clusters identified from the Millennium I+II runs as described in \citet{chiang13}. The minimum cluster mass is $10^{14} h^{-1} M_\odot$ at $z=0$. The comoving volume of the simulation is $(500/h)^3$~Mpc$^3$ or 0.364~Gpc$^3$. Our survey volume is estimated conservatively to be $2.13\times 10^6$~Mpc$^3$  assuming a flat redshift distribution at $z=3.4-4.2$ over a 28\arcmin$\times$28\arcmin\ area, which is 0.6\% of the Millennium  volume. 

We randomly pick a region matching our survey volume, and record the number of clusters therein and the position and the mass of each cluster.  The procedure is repeated 500,000 times. The median (mean) number of clusters is found to be 16.0 (16.5) with a standard deviation of 5.2; i.e., our survey volume is large enough to contain multiple clusters. 

If the LAE and photo-$z$ overdensities are part of a single very large structure, its combined mass would be enormous.  In \citet{dey16}, based on the level and extent of the LAE overdensity alone, we estimated  that the enclosed mass is $\gtrsim 10^{15}M_\odot$. As discussed in \S~\ref{overdensity},  the photo-$z$ overdensity should contain a comparable mass. Given that the two overdensities only partially overlap (and the regions of the peak overdensities do not overlap), a  conservative  limit on its combined mass is in the range of $(1.5-2.0)\times10^{15}M_\odot$.  We find the probability of these scenarios to be 4.4\% and 0.8\%, respectively. In the entire Millennium volume, eight and one  structures exist with masses above $1.5\times10^{15}M_\odot$ and $2\times10^{15}M_\odot$ respectively, corresponding to the comoving number density of $(2.2\pm0.8)\times10^{-8}$~Mpc$^{-3}$ and $(2.7\pm2.7)\times10^{-9}$~Mpc$^{-3}$, respectively. The most massive structure in the Millennium simulations has a total mass of  $2.4\times 10^{15}M_\odot$. The observational counterpart of such an ultramassive cluster may be the El Gordo system, which is a merging pair of Coma-analogs at $z=0.87$ \citep{Marriage11}. 

To test the possibility that the two overdensities are unrelated structures, we search for the cases in which   there are two Coma-like clusters (i.e., each with mass $\geq 10^{15}M_\odot$). This occurs only 3.6\% of the time. Finally, we assess how unlikely it is that the photo-$z$ overdensity lies at $z=3.72$ (see discussion in \S~\ref{sky_distribution}), which would put the distance between the two at 47~Mpc or 10~proper Mpc.  Only four distinct pairs of Coma analogs exist in the Millennium sample that are within 10~Mpc (physical) from each other. Two of those pairs have the physical separation of 5.2~Mpc and 5.3~Mpc from each other, and the other two at 9.3~Mpc and 9.8~Mpc. The separation for the latter is comparable to that between PC217.96+32.3 and the putative overdensity at $z=3.72$ \citep{lee18}. The likelihood of such a configuration falling into our survey is 0.8\%.  These considerations show that both scenarios are extremely unlikely to occur by chance, but also that it is not impossible. 

The overall density of the regions in and around PC217.96+32.3 is remarkably high.  Apart from the two overdensities we discuss here,  two other LAE overdensities lie within  $\sim 10$~Mpc (physical) north of of PC217.96+32.3 \citep{lee14}, one of which is spectroscopically confirmed and has  the estimated descendant mass of $\approx 6\times10^{14}M_\odot$ \citep{dey16}.  Given the distances between these system, it is unlikely they will coalesce into a single structure within the Hubble time, but rather, will evolve separately and  form structures similar to local superclusters \citep[e.g.,][]{einastoetal97,einasto14}.

\section{Summary}

Utilizing the multi-wavelength dataset taken in the sightline of PC217.96+32.3, a spectroscopically confirmed  protocluster at $z=3.78$, we have detected continuum-faint LAEs \citep{lee14,dey16}, UV-luminous  star-forming galaxies, candidates of passive galaxies and of young post-starburst galaxies with a strong Balmer/4000\AA\ break. Together, these constituent galaxies span  2--3 orders of magnitudes in both stellar masses and SFRs, highlighting diverse galaxy types residing in and around one of the largest structures discovered to date. Although we do not have spectroscopic redshifts for the new candidate protocluster members, the photometric redshift estimates suggest that they lie at or near the redshift of PC217.96+32.3. Based on our analyses, we conclude the following:\\

%\begin{enumerate}
\noindent [1] A significant overdensity $\delta_g\approx 7.8\pm 2.4 $ of massive star-forming galaxies is present in the field. 
The extent of the newly identified photo-$z$ overdensity only partially overlaps with that of the previously known and spectroscopically confirmed members, which are mostly LAEs; the two are offset by 3--4~Mpc in the east-west direction. 
While the origin of this separation is unclear, we speculate that the true extent of the structure may be larger than previously thought with a complex geometry only a part of which is traced by the LAE sample. This is presumably because the redshift of the main portion of the overdensity puts the Ly$\alpha$ emission just outside the bandpass of the narrow-band filter used for the LAE selection. If the combined overdensity traces a single structure, a conservative estimate would place its total mass  in the range of $(1.5-2.0)\times 10^{15}M_\odot$, making it a singularly large cosmic structure rarely seen  in  cosmological simulations.  However, we cannot rule out that the galaxy distributions are produced by a chance projection of two unassociated protoclusters at $z\sim4$, each of which will evolve into a Coma-like cluster by $z=0$. The likelihoods of both scenarios are extremely low ($<1$\%). \\

\noindent [2] We find a large excess ($\delta \Sigma_{\rm BBG}$$\approx$ 9--16) of ultramassive ($\gtrsim 10^{11}M_\odot$) galaxies exhibiting a strong Balmer/4000\AA\ break. Their SEDs are consistent with those of  passively evolving galaxies and of young post-starburst galaxies entering into quiescence. The sky distribution of BBGs appears to trace the full extent of the large scale structure rather than being concentrated in the highest density environments. We speculate that BBGs may represent the central and most massive inhabitants of the dark matter halos that are in the process of merging. Quenching of massive cluster ellipticals occurred in the epoch when the high-density environment did not adversely impact star formation activities therein, and long before these galaxies become part of a single coalesced structure. If confirmed, the presence of massive and quiescent galaxies as early as $z\sim 3.8$ would push back the formation epoch of the  cluster red sequence to  beyond $z_f\approx5$ in the largest clusters such as Coma. \\

\noindent [3] Stellar population parameters measured for all member candidates  span several orders of magnitude in the dynamic range of stellar masses, SFRs, and specific SFRs, showcasing the diverse constituents inhabiting the underlying large scale structure.  We find that the protocluster galaxies obey the same SFR-$M_{\rm star}$ scaling relation as the field galaxies. Our results suggest that the environmental effect on the stellar population properties of galaxy constituents is a subtle one at best. 
Alternatively, the impact of local environment manifests itself in producing extremely dust starburst systems, which would entirely elude our selection of galaxy candidates.  \\

%However, it is possible that local environment only impacts those residing in the highest density region, which produces extremely dusty starburst systems which would entirely elude our photo-z selection method. 
%More extensive spectroscopy and sensitive millimeter/far-IR imaging of the field are needed to further investigate such a possibility.\\

\noindent [4] While all galaxy types (LAEs, LBGs, and BBGs) show significant overdensities in the region, the BBGs show the largest overdensity.  
%The presence of a significant galaxy overdensity is evident in all galaxy types considered. However, the measured overdensities vary substantially.  
If they trace the same underlying structure, our results would be consistent with  the theoretical expectation that more massive galaxies are more biased tracers of the underlying matter. These results highlight the usefulness of using low-mass galaxies such as LAEs as the least biased visible tracers in quantifying the large-scale structures around massive protoclusters such as the one we have studied. Sensitive LAE surveys are therefore an efficient method to characterize large scale structure at high redshift, discover protoclusters, and as tracers of the physical processes responsible for cluster formation.

%\end{enumerate}
\acknowledgments
We thank the anonymous referee for helpful comments. KS thanks Denis Burgarella for the help on the usage of the CIGALE software, and Emiliano Merlin and Adriano Fontana for the help on using the TPHOT software. KSL thanks Yi-Kuan Chiang and Roderik Overzier for providing the Millennium Runs data for comparison, and Adam Muzzin and Thibaud Moutard for useful discussions and suggestions.
% KPNO
Based on observations at Kitt Peak National Observatory, National Optical Astronomy Observatory (NOAO Prop. IDs 2012A-0454, 2014A-0164, 2015A-0168, 2016A-0185; PI: K.-S. Lee), which is operated by the Association of Universities for Research in Astronomy (AURA) under cooperative agreement with the National Science Foundation. The authors are honored to be permitted to conduct astronomical research on Iolkam Du'ag (Kitt Peak), a mountain with particular significance to the Tohono O'odham. 
% Keck 
%Data presented herein were obtained at the W. M. Keck Observatory using telescope time allocated to the National Aeronautics and Space Administration through the agency's scientific partnership with the California Institute of Technology and the University of California. 
%The Observatory was made possible by the generous financial support of the W. M. Keck Foundation. The authors wish to recognize and acknowledge the very significant cultural role and reverence that the summit of Mauna Kea has always had within the indigenous Hawaiian community. We are most privileged to be able to conduct observations from this mountain. 
This work was supported by a NASA Keck PI Data Award, administered by the NASA Exoplanet Science Institute. We thank NASA for support, through grants NASA/JPL\# 1497290 and 1520350. 
%Subaru
This paper is based in part on data collected at Subaru Telescope, which is operated by the National Astronomical Observatory of Japan. 
%Spitzer
This work is based in part on observations made with the Spitzer Space Telescope, which is operated by the Jet Propulsion Laboratory, California Institute of Technology under a contract with NASA.
% 2mass 
This publication makes use of data products from the Two Micron All Sky Survey, which is a joint project of the University of Massachusetts and the Infrared Processing and Analysis Center/California Institute of Technology, funded by the National Aeronautics and Space Administration and the National Science Foundation.
% AD
AD's research was supported in part by the National Optical Astronomy Observatory (NOAO), which is operated by the Association of Universities for Research in Astronomy (AURA), Inc. under a cooperative agreement with the National Science Foundation. This work was performed in part at Aspen Center for Physics, which is supported by National Science Foundation grant PHY-1607611.

\bibliographystyle{apj}
\bibliography{myrefs}



\begin{sidewaystable}\label{bbg_table}
\centering
\begin{threeparttable}
\caption{Catalog of BBG candidates}
\begin{tabular}{ccccccccccccc}
\hline
ID & R.A. (J2000) & Decl. (J2000) & $B_W$ & $R$ & $I$ & $y$ & $H$ & $K_S$ & [3.6] & [4.5] & [5.8] & [8.0] \\
\hline

Q112 & 217.953683 & 32.28450 & $>$27.41 &  $>$27.02 & $>$26.67 & $>$26.34 & 24.53$\pm$0.29 &  23.21$\pm$0.10 & 21.98$\pm$0.15 & 21.56$\pm$0.14 & $>$21.23 & $>$20.90    \\
Q2396  & 217.741855 & 32.32550  & $>$27.41  & 26.36$\pm$0.37   & 25.89$\pm$0.47   & 25.47$\pm$0.37   & 24.27$\pm$0.23   & 23.00$\pm$0.14    & 22.96$\pm$0.26       & 22.56$\pm$0.22       & $>$21.23      & $>$20.90         \\
Q3161  & 217.721483 & 32.19860  & $>$27.41 & $>$27.02   & $>$26.67  & $>$26.34   & 24.25$\pm$0.24   & 22.89$\pm$0.10     & 21.83$\pm$0.20        & 21.33$\pm$0.14       & --   & --           \\
Q3268  & 217.943585 & 32.34119 & $>$27.41    & 26.45$\pm$0.48   & $>$26.67  & 25.98$\pm$0.76   & 24.10$\pm$0.32   & 22.70$\pm$0.14    & 22.15$\pm$0.20       & 21.68$\pm$0.18       & $>$21.23      & $>$20.90           \\
Q3296  & 217.931437 & 32.20151 & $>$27.41    & $>$27.02   & $>$26.67    & 24.78$\pm$0.46   & 24.06$\pm$0.33   & 22.69$\pm$0.13    & 22.51$\pm$0.63       & 22.78$\pm$0.53       & --    & $>$20.90        \\
P3374  & 217.766615 & 32.34254 & $>$27.41    & 25.37$\pm$0.16   & 25.19$\pm$0.25   & 24.98$\pm$0.27   & 23.68$\pm$0.15   & 22.37$\pm$0.08    & 22.26$\pm$0.14       & 21.93$\pm$0.12       & 20.84 $\pm$0.35       & 20.22$\pm$0.16           \\
P3433  & 218.098397 & 32.20417 & $>$27.41     & 26.10$\pm$0.28   & 25.57$\pm$0.31   & 26.01$\pm$0.73   & 24.72$\pm$0.28   & 23.32$\pm$0.12    & 22.91$\pm$0.56       & 22.61$\pm$0.32       & --   & $>$20.90          \\
Q4670  & 217.914775 & 32.22440  & $>$27.41    & $>$27.02   & $>$26.67  & $>$26.34   & 23.89$\pm$0.23   & 22.63$\pm$0.08    & 21.52$\pm$0.18       & 21.43$\pm$0.15       & -- & $>$20.90          \\
Q4943  & 218.106545 & 32.37317 & $>$27.41  & 26.55$\pm$0.38   & $>$26.67 & 25.55$\pm$0.41   & 24.88$\pm$0.40    & 23.29$\pm$0.22    & 22.68$\pm$0.27       & 22.31$\pm$0.26       & $>$21.23      & $>$20.90           \\
Q4948  & 218.027168 & 32.22880  & $>$27.41 & $>$27.02  & $>$26.67   & $>$26.34    & 24.63$\pm$0.29   & 23.28$\pm$0.15    & 22.75$\pm$0.53       & 22.74$\pm$0.36       & -- & $>$ 20.90           \\
P5479  & 218.107794 & 32.38329 & $>$27.41    & 24.91$\pm$0.21   & 24.94$\pm$0.24   & 24.85$\pm$0.27   & 24.61$\pm$0.35   & 23.14$\pm$0.17    & 22.92$\pm$0.27       & 22.42$\pm$0.24       & $>$21.23       & $>$20.90         \\
Q5595  & 217.762752 & 32.38484 & $>$27.41   & $>$27.02  & $>$26.67 & 25.67$\pm$0.53   & 23.86$\pm$0.18   & 22.55$\pm$0.08    & 21.69$\pm$0.12       & 21.43$\pm$0.11       & 21.05 $\pm$0.33       & 20.78$\pm$0.28          \\
Q5788  & 217.833271 & 32.38915 & $>$27.41    & 26.49$\pm$0.29   & 26.35$\pm$0.49   & $>$26.34   & 24.62$\pm$0.26   & 23.19$\pm$0.11    & 22.96$\pm$0.23       & 22.65$\pm$0.21       & $>$21.23      & $>$20.90         \\
Q6407  & 217.756014 & 32.40108 & $>$27.41    & $>$27.02   & 25.85$\pm$0.53   & 26.08$\pm$0.86   & $>$25.05   & 22.93$\pm$0.15    & 22.90$\pm$0.37       & 22.64$\pm$0.33       & $>$21.23        & $>$20.90          \\
Q6497  & 217.897260  & 32.25346 & $>$27.41    & 26.61$\pm$0.78   & 25.85$\pm$0.69   & 24.97$\pm$0.50    & 23.98$\pm$0.29   & 22.68$\pm$0.11    & 21.84$\pm$0.36       & 22.16$\pm$0.39       & --     & $>$20.90           \\
Q6936  & 217.732113 & 32.41123 & $>$27.41     & $>$27.02   & $>$26.67  & $>$26.34   & $>$25.05   & 23.76$\pm$0.22    & 22.99$\pm$0.29       & 22.57$\pm$0.26       & $>$21.23       & $>$20.90           \\
Q7103  & 217.884829 & 32.41509 & $>$27.41     & $>$27.02  & $>$26.67  & $>$26.34   & $>$25.05   & 23.68$\pm$0.19    & 23.18$\pm$0.39       & 22.99$\pm$0.35       & 20.47$\pm$0.37       & $>$20.90            \\
Q7163  & 217.902270  & 32.41554 & $>$27.41    & $>$27.02   & $>$26.67  & 25.31$\pm$0.50    & $>$25.05   & 22.66$\pm$0.13    & 22.99$\pm$0.44       & 22.80$\pm$0.39       & 21.19$\pm$0.77       & $>$20.90           \\
Q7361  & 218.138166 & 32.41936 & $>$27.41    & $>$27.02   & $>$26.67  & 26.15$\pm$0.70    & 24.39$\pm$0.36   & 23.05$\pm$0.14    & 21.68$\pm$0.18       & 21.25$\pm$0.18       & $>$21.23      & 20.45$\pm$0.28           \\
Q7775  & 218.095899 & 32.42599 & $>$27.41  & $>$27.02   & $>$26.67  & $>$26.34  & 23.99$\pm$0.19   & 22.79$\pm$0.12    & 21.53$\pm$0.14       & 21.13$\pm$0.14       & 20.51  $\pm$0.24       & $>$20.90            \\
Q7961  & 217.658322 & 32.42931 & $>$27.41  & 26.61$\pm$0.45   & 26.22$\pm$0.57   & 25.72$\pm$0.54   & 24.54$\pm$0.27   & 23.26$\pm$0.18    & 22.66$\pm$0.28       & 22.70$\pm$0.29       & $>$21.23       & 20.53$\pm$0.33           \\
Q8038  & 217.712417 & 32.43066 & $>$27.41    & $>$27.02  & $>$26.67  & $>$26.34 & 24.22$\pm$0.28   & 22.77$\pm$0.13    & 22.15$\pm$0.19       & 21.69$\pm$0.18       & 20.36 $\pm$0.27       & $>$20.90            \\
Q8068  & 217.871038 & 32.43181 & $>$27.41     & $>$27.02   & $>$26.67   & 26.12$\pm$1.07   & 24.33$\pm$0.27   & 22.75$\pm$0.16    & 21.72$\pm$0.24       & 21.39$\pm$0.23       & 20.27$\pm$0.44       & $>$20.90           \\
Q8108  & 217.737275 & 32.27759 & $>$27.41    & $>$27.02   & $>$26.67   & 25.70$\pm$0.57   & 24.23$\pm$0.27   & 22.93$\pm$0.15    & 22.43$\pm$0.24       & 22.03$\pm$0.19       & $>$21.23       & $>$20.90            \\
Q8847  & 217.978870  & 32.44496 & $>$27.41    & $>$27.02   & $>$26.67  & $>$26.34   & 24.25$\pm$0.25   & 22.97$\pm$0.16    & 22.10$\pm$0.22       & 21.79$\pm$0.21       & 20.22 $\pm$0.28       & $>$20.90           \\
Q9512  & 217.951934 & 32.30146 & $>$27.41    & $>$27.02    & 26.02$\pm$0.54   & 24.70$\pm$0.26   & 23.45$\pm$0.12   & 22.24$\pm$0.06    & 21.47$\pm$0.10        & 21.56$\pm$0.10        & $>$21.23      & $>$20.90            \\
P9576  & 217.865190  & 32.45845 & $>$27.41     & 25.87$\pm$0.18   & 25.81$\pm$0.35   & 25.23$\pm$0.26   & 24.00$\pm$0.15   & 22.73$\pm$0.09    & 23.09$\pm$0.21       & 22.90$\pm$0.20        & 20.87$\pm$0.25       & $>$20.90          \\
Q10133 & 217.948664 & 32.46798 & $>$27.41    & $>$27.02  & $>$26.67  & 25.82$\pm$0.55   & 24.12$\pm$0.24   & 22.79$\pm$0.10     & 21.87$\pm$0.15       & 21.51$\pm$0.14       & $>$21.23      & $>$20.90          \\
Q10226 & 217.807902 & 32.47021 & $>$27.41 & $>$27.02   & $>$26.67  & $>$26.34   & 24.35$\pm$0.23   & 22.88$\pm$0.11    & 22.96$\pm$0.23       & 22.72$\pm$0.22       & $>$21.23      & $>$20.90           \\
P10316 & 217.867090  & 32.47131 & 26.79$\pm$0.30    & 25.61$\pm$0.19   & 25.28$\pm$0.27   & 25.01$\pm$0.28   & 23.62$\pm$0.15   & 22.37$\pm$0.08    & 22.01$\pm$0.12       & 21.77$\pm$0.12       & 20.58$\pm$0.22       & $>$20.90          \\
Q10386 & 218.147692 & 32.47255 & $>$27.41   & $>$27.02  & $>$26.67  & $>$26.34   & 23.35$\pm$0.27   & 22.08$\pm$0.13    & 20.63$\pm$0.16       & 20.39$\pm$0.16       & 20.69  $\pm$0.52       & $>$20.90         \\
Q10794 & 217.902578 & 32.48004 & $>$27.41   & $>$27.02   & $>$26.67 & $>$26.34   & $>$25.05    & 23.71$\pm$0.17    & 23.17$\pm$0.43       & 22.72$\pm$0.34       & 20.95$\pm$0.55       & $>$20.90          \\
Q12526 & 217.897738 & 32.50942 & $>$27.41   & $>$27.02   & $>$26.67 & 25.98$\pm$0.60    & 24.65$\pm$0.48   & 23.34$\pm$0.18    & 22.39$\pm$0.24       & 22.37$\pm$0.27       & $>$21.23      & 20.80$\pm$0.52            \\
Q12773 & 218.051315 & 32.51337 & $>$27.41    & $>$27.02  & $>$26.67  & $>$26.34   & 24.04$\pm$0.18   & 22.83$\pm$0.09    & 21.82$\pm$0.13       & 21.49$\pm$0.12       & $>$21.23      & 20.55$\pm$0.21           \\
Q13609 & 217.935053 & 32.52815 & $>$27.41    & $>$27.02   & 26.60$\pm$0.66   & 25.80$\pm$0.46   & 24.84$\pm$0.34   & 23.58$\pm$0.12    & 23.26$\pm$0.30        & 22.90 $\pm$0.26       & $>$21.23      & $>$20.90         \\
P15053 & 218.146435 & 32.55216 & $>$27.41   & 25.73$\pm$0.32   & 25.25$\pm$0.36   & 24.47$\pm$0.31   & 24.27$\pm$0.37   & 22.92$\pm$0.23    & 22.34$\pm$0.28       & 22.35$\pm$0.31       & $>$21.23      & $>$20.90          \\
Q15256 & 217.728406 & 32.55562 & $>$27.41    & $>$27.02   & $>$26.67  & $>$26.34  & 24.35$\pm$0.29   & 22.90$\pm$0.14    & 22.50$\pm$0.23       & 22.12$\pm$0.21       & $>$21.23      & $>$20.90          \\
P15333 & 217.810823 & 32.55629 & 27.10$\pm$0.60     & 24.83 $\pm$0.22   & 24.57$\pm$0.27   & 24.13$\pm$0.26   & 23.84$\pm$0.31   & 22.59$\pm$0.16    & 21.93$\pm$0.23       & 21.76$\pm$0.23       & $>$21.23      & $>$20.90           \\
Q17001 & 218.005890  & 32.58427 & $>$27.41   & $>$27.02   & $>$26.67  & $>$26.34   & 24.65$\pm$0.31   & 23.19$\pm$0.11    & 22.27$\pm$0.17       & 21.94$\pm$0.17       & $>$21.23         & $>$20.90           \\
Q17800 & 217.704266 & 32.59865 & $>$27.41   & $>$27.02  & 26.59$\pm$0.84   & 26.20$\pm$0.81   & 24.59$\pm$0.29   & 23.11$\pm$0.12    & 22.08$\pm$0.18       & 21.64$\pm$0.17       & $>$21.23       & $>$20.90            \\
Q17832 & 217.993321 & 32.59982 & $>$27.41     & $>$27.02  & 25.80$\pm$0.41   & 25.04$\pm$0.30    & 24.25$\pm$0.29   & 22.77$\pm$0.12    & 22.53$\pm$0.22       & 22.18$\pm$0.20        & $>$21.23       & $>$20.90          \\
Q17976 & 218.067814 & 32.60220  & $>$27.41    & $>$27.02   & $>$26.67 & 25.85$\pm$0.53   & 24.34$\pm$0.26   & 23.09$\pm$0.13    & 22.19$\pm$0.18       & 22.23$\pm$0.19       & $>$21.23      & $>$20.90            \\
Q18587 & 218.051648 & 32.61501 & $>$27.41  & $>$27.02   & $>$26.67 & $>$26.34   & 24.07$\pm$0.31   & 22.84$\pm$0.14    & 21.66$\pm$0.21       & 21.30$\pm$0.20        & $>$21.23      & $>$20.90          \\
\hline

\end{tabular}

    \begin{tablenotes}
      \small
      \item Notes: The letter in front of the ID number represents the type of BBGs: `P' for post-starbursts and `Q' for quiescent galaxies. In the case of non-detection, the corresponding 2$\sigma$ limiting magnitude is given. The sign `--' is used when a source is out of image bounds or lies in the region with less than 20\% of the maximum exposure time. 
    \end{tablenotes}

\end{threeparttable}
\end{sidewaystable}

\end{document}